\documentclass[onecolumn,12pt,journal,draftclsnofoot]{IEEEtran}
\IEEEoverridecommandlockouts
\usepackage{epsfig,epstopdf,setspace,subfigure}
\usepackage{amsmath,amssymb,cite}
\usepackage{enumerate}
\newtheorem{proposition}{Proposition}

\newtheorem{theorem}{Theorem}
\newtheorem{definition}{Definition}

\begin{document}
\title{A Game-Theoretic View of the Interference Channel: Impact of Coordination and Bargaining}
\author{Xi~Liu,~\IEEEmembership{Student~Member,~IEEE,}
        and Elza Erkip,~\IEEEmembership{Senior Member,~IEEE}
\thanks{This material is based upon work partially supported by NSF Grant No.
0635177, and by the Center for Advanced Technology in Telecommunications
(CATT) of Polytechnic Institute of NYU.}
}
\maketitle
\doublespacing
\begin{abstract}
This work considers coordination and bargaining between two selfish users over a Gaussian interference channel. The usual information theoretic approach assumes full cooperation among users for codebook and rate selection. In the scenario investigated here, each user is willing to coordinate its actions only when an incentive exists and benefits of cooperation are fairly allocated. The users are first allowed to negotiate for the use of a simple Han-Kobayashi type scheme with fixed power split. Conditions for which users have incentives to cooperate are identified. Then, two different approaches are used to solve the associated bargaining problem. First, the Nash Bargaining Solution (NBS) is used as a tool to get fair information rates and the operating point is obtained as a result of an optimization problem. Next, a dynamic alternating-offer bargaining game (AOBG) from bargaining theory is introduced to model the bargaining process and the rates resulting from negotiation are characterized. The relationship between the NBS and the equilibrium outcome of the AOBG is studied and factors that may affect the bargaining outcome are discussed. Finally, under certain high signal-to-noise ratio regimes, the bargaining problem for the generalized degrees of freedom is studied. 
\end{abstract}

\section{Introduction}

Interference channel (IC) is a fundamental model in information theory for studying interference in communication systems. In this model, multiple senders transmit independent messages to their corresponding receivers via a common channel. The capacity region or the sum-rate capacity for the two-user Gaussian IC is only known in special cases such as the strong interference case \cite{refereces:Sato81}\cite{references:Han81} or the noisy interference case\cite{references:Shang09}; the characterization of the capacity region for the general case remains an open problem. Recently, it has been shown in \cite{references:Etkin08} that a simplified version of a scheme due to Han and Kobayashi \cite{references:Han81} results in an achievable rate region that is within one bit of the capacity region of the complex Gaussian IC for all values of channel parameters. However, any type of Han-Kobayashi (H-K) scheme requires full cooperation\footnote{Throughout the paper, ``cooperation'' means cooperation for the choice of transmission strategy including codebook and rate selection, which is different from cooperation in information transmission as in cooperative communications\cite{references:Sendonaris03}.} between the two users through the choice of transmission strategy. In practice, users are selfish in the sense that they choose a transmission strategy to maximize their own rates. They may not have an incentive to comply with a certain rule as in the H-K scheme and therefore not all rate pairs in an achievable rate region are actually attainable. When there is no coordination among the users, interference is usually treated as noise, which is information theoretically suboptimal in most cases.

In this paper, we study a scenario where two users operating over a Gaussian IC are selfish but willing to coordinate and bargain to get fair information rates. When users have conflicting interests, the problem of achieving efficiency and fairness could be formulated as a game-theoretic problem. The Gaussian IC was studied using noncooperative game theory in \cite{references:Yu00}\cite{references:Etkin07}\cite{references:Larsson08}, where it was assumed that the receivers treat the interference as Gaussian noise. For the related Gaussian multiple-access channel (MAC), it was shown in \cite{references:Gajic08} that in a noncooperative rate game with two selfish users choosing their transmission rates independently, all points on the dominant face of the capacity region are pure strategy Nash Equilibria (NE). However, no single NE is superior to the others, making it impossible to single out one particular NE to operate at. The authors resorted to a mixed strategy which is inefficient in performance. Noncooperative information theoretic games were considered by Berry and Tse in \cite{references:Berry08} assuming that each user can select any encoding and decoding strategy to maximize its own rate and a Nash equilibrium region was characterized for a class of deterministic IC's. Extensions were made to a symmetric Gaussian IC in \cite{references:Berry09}.

Another game theoretic approach for studying interfering links is through cooperative game theory. Coalitional games were studied in \cite{references:La04} for a Gaussian MAC and in \cite{references:Mathur_Sankar_Mandayam06}\cite{references:Mathur_Sankar_Mandayam08} for Gaussian IC's. In \cite{references:Mathur_Sankar_Mandayam06}, the Nash Bargaining Solution (NBS) is considered for a Gaussian IC under the assumption of receiver cooperation, effectively translating the channel to a MAC. In \cite{references:Han05}, the NBS was used as a tool to develop a fair resource allocation algorithm for uplink multi-user OFDMA systems. References \cite{references:Leshem06}\cite{references:Leshem08} analyzed the NBS for the flat and frequency selective fading IC under the assumption of time or frequency division multiplexing (TDM/FDM). The emphasis there was on the weak interference case\footnote{In the NBS discussed in \cite{references:Leshem06}\cite{references:Leshem08}, it is assumed that a unique NE of an Gaussian interference game defined there exists and is selected as disagreement point. Typically, the NE is unique only when both interferences are weaker than the desired signals.}. However, as we will show later, for the strong and mixed interference regimes, the NBS based on TDM/FDM may not perform very well, due to the suboptimality of TDM/FDM in those regimes. Game theoretic solutions for the MISO and MIMO IC based on bargaining have been investigated in \cite{references:Jorswieck}\cite{references:Larsson08}\cite{references:zengmao09}, where two or more users negotiate for an agreement on the choice of beamforming vectors or source covariance matrices whereas single-user detection is employed at the receivers.

In this paper, unlike the above literature, we allow for the use of H-K type schemes thereby resulting in a larger rate region and let the two users bargain on choices of codebook and rate to improve their achieved rates or generalized degrees of freedom compared with the uncoordinated case. We propose a two-phase mechanism for coordination between users. In the first phase, the two users negotiate and only if certain incentive conditions are satisfied they agree to use a simple H-K type scheme with a fixed power split that gives the optimal or close to optimal set of achievable rates\cite{references:Etkin08}. For different types of IC's, we study the incentive conditions for users to coordinate their transmissions. In the second phase, provided that negotiation in the first phase is successful, the users bargain for rates over the H-K achievable rate region to find an acceptable operating point. Our primary contribution is the application of two different bargaining ideas from game theory to address the bargaining problem in the second phase: the cooperative bargaining approach using NBS and the noncooperative bargaining approach using alternating-offer bargaining games (AOBG). The advantage of the NBS is that it not only provides a Pareto optimal operating point from the point of view of the entire system, but is also consistent with the fairness axioms of game theory. However, one of the assumptions upon which cooperative bargaining is built is that the users are committed to the agreement reached in bargaining when the time comes for it to be implemented \cite{references:Binmore98}. In this sense, the NBS may not necessarily be the agreement reached in practice. Before the NBS can be used as the operating point, some form of centralized coordination is still needed to ensure that all the parties involved jointly agree to operate at such a point. In an unregulated environment, a centralized authority may be lacking and in such cases more realistic bargaining between users through communication over a side channel may become necessary. Besides, in most works that designate the NBS as a desired solution, each user's cost of delay in bargaining is not taken into account and little is known regarding how bargaining proceeds. Motivated by all these, we will also study the bargaining problem under the noncooperative bargaining model AOBG \cite{references:Binmore98}\cite{references:Martin} over the IC. This approach is different from the NBS in that it models the bargaining process between users explicitly as a non-cooperative multi-stage game in which the users alternate making offers until one is accepted. The equilibrium of such a game describes what bargaining strategies would be adopted by the users and thus provides a nice prediction to the result of noncooperative bargaining.  To the best of our knowledge, our work provides the first application of dynamic AOBG from bargaining theory to network information theory.

Under the cooperative bargaining approach, the computation of the NBS over the H-K rate region is formulated as a convex optimization problem. Results show that the NBS exhibits significant rate improvements for both users compared with the uncoordinated case. Under the noncooperative bargaining approach, the two-user IC bargaining problem is considered in an uncoordinated environment where the ongoing bargaining may be interrupted, for example, by other users wishing to access the channel. Each user's cost of delay in bargaining is derived from an exogenous probability which characterizes the risk of breakdown of bargaining due to some outside intervention. The AOBG with risk of breakdown is introduced to model the bargaining process and the negotiation outcome in terms of achievable rates is analyzed. We show that the equilibrium outcome of the AOBG lies on the individual rational efficient frontier of the rate region with its exact location depending on the exogenous probabilities of breakdown. When the breakdown probabilities are very small, it is shown that the equilibrium outcome approaches the Nash solution.

The remainder of this paper is organized as follows. In Section II, we present the channel model, describe the achievable region of a simple H-K type scheme using Gaussian codebooks and review the concept of the NBS and that of AOBG from game theory. We first illustrate how two selfish users bargain over the Gaussian MAC to get higher rates for both in Section III and then present the mechanism of coordination and bargaining for the two users over the Gaussian IC in Section IV. In Section V we consider the bargaining problem in certain high SNR regimes when the utility of each selfish user is measured by achieved generalized degree of freedom (g.d.o.f.) instead of allocated rate, and finally we draw conclusions in Section VI.

Before, we proceed to the next section, we introduce some notations that will be used in this paper.
\begin{itemize}
\item Italic letters (e.g. $x$, $X$) denote scalars; and bold letters $\mathbf{x}$ and $\mathbf{X}$ denote column vectors or matrices.
\item $\mathbf{0}$ denotes the all-zero vector.
\item $\mathbf{X}^t$ and $\mathbf{X}^{-1}$ denote the transpose and inverse of the matrix $\mathbf{X}$ respectively.
\item For any two vectors $\mathbf{u}$ and $\mathbf{v}$, we denote $\mathbf{u} \geq \mathbf{v}$ if and only if $u_i \geq v_i$ for all $i$. $\mathbf{u} \leq \mathbf{v}$, $\mathbf{u} > \mathbf{v}$ and $\mathbf{u} <\mathbf{v}$ are defined similarly.
\item $C(\cdot)$ is defined as $C(x) = \frac{1}{2}\log_2(1+x)$.
\item  $(\cdot)^+$ means $\max(\cdot,0)$.
\item $\mathbb{R}$ denotes the set of real numbers.
\end{itemize}

\section{System Model and Preliminaries}
\subsection{Channel Model}
In this paper, we focus on the two-user standard Gaussian IC \cite{refereces:Sason04} as shown in Fig. 1
\begin{eqnarray}
Y_{1,t} = X_{1,t} + \sqrt{a}X_{2,t}+Z_{1,t}\\
Y_{2,t} = \sqrt{b}X_{1,t} + X_{2,t} + Z_{2,t}
\end{eqnarray}
where $X_{i,t}$ and $Y_{i,t}$, $t = 1,...,n$ represent the input and output at transmitter and receiver $i \in \{1,2\}$ at time $t$, respectively, and $Z_{1,t}$ and $Z_{2,t}$ are i.i.d. Gaussian with zero mean and unit variance. Receiver $i$ is only interested in the message sent by transmitter $i$. For a given block length $n$, user $i$ sends a message $W_i \in \{1,2,..,2^{nR_i}\}$ by encoding it to a codeword $\mathbf{X_i} = (X_{i,1},X_{i,2},...,X_{i,n})$. The codewords $\mathbf{X_1}$ and $\mathbf{X_2}$ satisfy the average power constraints given by
\begin{equation}
\frac{1}{n}\sum_{t=1}^n X_{1,t}^2 \leq P_1, \quad
\frac{1}{n}\sum_{t=1}^n X_{2,t}^2 \leq P_2 \label{eqn:powerconst}
\end{equation}
Receiver $i$ observes the channel output $\mathbf{Y_i} = (Y_{i,1},Y_{i,2},...,Y_{i,n})$ and uses a decoding function $f_i: \mathbb{R}^n \rightarrow \{1,..,2^{nR_i}\}$ to get the estimate $\hat{W}_i$ of the transmitted message $W_i$. We define the average probabilities of error by the expressions
\begin{eqnarray}
p_{e,1}^n = \text{P}\{f_1(\mathbf{Y_1})\neq W_1\}\\
p_{e,2}^n = \text{P}\{f_2(\mathbf{Y_2})\neq W_2\}
\end{eqnarray}
and
\begin{equation}
p_{e}^n = \max \{p_{e,1}^n,p_{e,2}^n\}.
\end{equation}
A rate pair $(R_1, R_2)$ is said to be achievable if there is a sequence of $(2^{nR_1},2^{nR_2},n)$ codes with $p_e^n\rightarrow 0$ as $n\rightarrow \infty$. The capacity region of the interference channel is the closure of the set of all achievable rate pairs.

Constants $\sqrt{a}$ and $\sqrt{b}$ represent the real-valued channel gains of the interfering links. Depending on the values of $a$ and $b$, the two-user Gaussian IC can be classified as strong, weak and mixed. If $a \geq 1$ and $b \geq 1$, the channel is {\em strong} Gaussian IC; if $0<a<1$ and $0<b<1$, the channel is {\em weak} Gaussian IC; if either $0<a<1$ and $b\geq 1$, or $0<b<1$ and $a\geq 1$, the channel is {\em mixed} Gaussian IC. We let $\text{SNR}_i = P_i$ be the signal to noise ratio (SNR) of user $i$, and $\text{INR}_1 = aP_2(\text{INR}_2 = bP_1)$ be the interference to noise ratio (INR) of user 1(2).  

\subsection{The Han-Kobayashi Rate Region}
The best known inner bound for the two-user Gaussian IC is given by the full H-K achievable region \cite{references:Han81}. Even when the input distributions in the H-K scheme are restricted to be Gaussian, computation of the full H-K region by taking the union of all power splits into common and private messages and time sharing remains difficult due to numerous degrees of freedom involved in the problem \cite{references:Khandani09}. Therefore for the purpose of evaluating and computing bargaining solutions, we assume users employ Gaussian codebooks with equal length codewords and consider a simplified H-K type scheme with fixed power split and no time-sharing as in \cite{references:Etkin08}. Let $\alpha \in [0,1]$ and $\beta \in [0,1]$ denote the fractions of power allocated to the private messages (messages only to be decoded at intended receivers) of user 1 and user 2 respectively. 
We define $\mathcal{F}$ as the collection of all rate pairs $(R_1,R_2)\in \mathbb{R}^2_{+}$ satisfying
\begin{equation}
R_1 \leq \phi_1 = C\left(\frac{P_1}{1+a\beta P_2}\right)\label{eqn:reg1}
\end{equation}
\begin{equation}
R_2 \leq \phi_2 = C\left(\frac{P_2}{1+b\alpha P_1}\right)\label{eqn:reg2}
\end{equation}
\begin{equation}
R_1 + R_2 \leq \phi_3 = \min\{\phi_{31},\phi_{32},\phi_{33}\}\label{eqn:reg3}
\end{equation}
with
\begin{equation}
\phi_{31} = C\left(\frac{P_1+a(1-\beta)P_2}{1+a\beta P_2}\right) + C\left(\frac{\beta P_2}{1+b\alpha P_1}\right)
\end{equation}
\begin{equation}
\phi_{32} = C\left(\frac{\alpha P_1}{1+a\beta P_2}\right) + C\left(\frac{P_2+b(1-\alpha)P_1}{1+b\alpha P_1}\right)
\end{equation}
\begin{equation}
\phi_{33} = C\left(\frac{\alpha P_1+a(1-\beta)P_2}{1+a\beta P_2}\right) + C\left(\frac{\beta P_2+b(1-\alpha)P_1}{1+b\alpha P_1}\right)
\end{equation}
and
\begin{equation}
\begin{array}{l l}
2R_1+R_2 \leq \phi_4  = &\displaystyle C\left(\frac{P_1+a(1-\beta)P_2}{1+a\beta P_2}\right) + C\left(\frac{\alpha P_1}{1+a\beta P_2}\right)\\
&\displaystyle + C\left(\frac{\beta P_2+b(1-\alpha)P_1}{1+b\alpha P_1}\right)
\end{array}\label{eqn:reg4}
\end{equation}
\begin{equation}
\begin{array}{l l}
R_1+2R_2 \leq \phi_5  = &\displaystyle C\left(\frac{P_2+b(1-\alpha)P_1}{1+b\alpha P_1}\right) + C\left(\frac{\beta P_2}{1+b\alpha P_1}\right)\\
&\displaystyle + C\left(\frac{\alpha P_1+a(1-\beta)P_2}{1+a\beta P_2}\right)
\end{array}\label{eqn:reg5}
\end{equation}

The region $\mathcal{F}$ is a polytope and a function of $\alpha$ and $\beta$. We denote the H-K scheme that achieves the rate region $\mathcal{F}$ by $\text{HK}(\alpha,\beta)$. For convenience, we also represent $\mathcal{F}$ in a matrix form as $\mathcal{F} = \{\mathbf{R}|\mathbf{R} \geq \mathbf{0},\: \mathbf{R} \leq \mathbf{R}^1,\: \text{and}\:\mathbf{A}_0\mathbf{R} \leq \mathbf{B}_0\}$, where $\mathbf{R} = (R_1\:R_2)^t$, $\mathbf{R}^1 = (\phi_1\:\phi_2)^t$, $\mathbf{B}_0 = (\phi_3\:\phi_4\:\phi_5)^t$, and
\begin{equation}
\mathbf{A}_0 = \left(
\begin{array}{c c c}
1 & 2 & 1\\
 1 & 1 & 2
\end{array}
\right)^t
\end{equation}

In the strong interference regime $a\geq1$ and $b\geq1$, the capacity region is known \cite{refereces:Sato81}\cite{references:Han81} and is achieved by $\text{HK}(0,0)$, i.e., both users send common messages only to be decoded at both destinations. This capacity region is the collection of all rate pairs $(R_1,R_2)$ satisfying
\begin{align}
&R_1 \leq C(P_1),\nonumber\\
&R_2 \leq C(P_2),\nonumber\\
&R_1 + R_2 \leq \phi_6 = \min\{C(P_1+aP_2), C(bP_1+P_2)\}\label{eqn:cap_strong}
\end{align}
Note that $\phi_6 = \phi_3$ for $\alpha = \beta = 0$.

\subsection{Overview of Bargaining Games}


A two-player bargaining problem consists of a pair $(\mathcal{G},\mathbf{g}^0)$ where $\mathcal{G}$ is a closed convex subset of $\mathbb{R}^2$, $\mathbf{g}^0 = (g_1^0\; g_2^0)^t$
is a vector in $\mathbb{R}^2$, and the set $\mathcal{G}\cap \{\mathbf{g}|\mathbf{g}\geq \mathbf{g}^0\}$ is nonempty
and bounded. Here $\mathcal{G}$ is the set of all possible payoff allocations or agreements that the two players can jointly achieve, and $\mathbf{g}^0 \in \mathcal{G}$ is the payoff allocation that results if players fail to agree. We refer to $\mathcal{G}$ as the \emph{feasible set} and to $\mathbf{g}^0$ as the \emph{disagreement point}. The set $\mathcal{G}\cap \{\mathbf{g}|\mathbf{g}\geq \mathbf{g}^0\}$ is a subset of $\mathcal{G}$ which contains all payoff allocations no worse than $\mathbf{g}^0$. We refer to it as the \emph{individual rational feasible set}. We say the bargaining problem $(\mathcal{G},\mathbf{g}^0)$ is {\em essential} iff there exists at least one
allocation $\mathbf{g}'$ in $\mathcal{G}$ that is strictly better for both players than $\mathbf{g}^0$, i.e., the set
$\mathcal{G} \cap \{\mathbf{g}|\mathbf{g}>\mathbf{g}^0\}$ is nonempty; we say $(\mathcal{G},\mathbf{g}^0)$
is \emph{regular} iff $\mathcal{G}$ is essential and for any payoff allocation $\mathbf{g}$ in $\mathcal{G}$,
\begin{equation}
\text{if } g_1>g_1^0, \text{ then } \exists \check{\mathbf{g}} \in \mathcal{G} \text{ such that } g_1^0\leq \check{g}_1<g_1 \text{ and } \check{g}_2>g_2,\label{eqn:regular1}
\end{equation}
\begin{equation}
\text{if } g_2>g_2^0, \text{ then } \exists \hat{\mathbf{g}} \in \mathcal{G} \text{ such that } g_2^0\leq \hat{g}_2<g_2 \text{ and } \hat{g}_1>g_1,\label{eqn:regular2}
\end{equation}
Here (\ref{eqn:regular1}) and (\ref{eqn:regular2}) state that whenever a player gets strictly higher payoff than in the disagreement point, then there exists another allocation such that the payoff of the player is reduced while the other player's payoff is strictly increased.

An agreement $\mathbf{g}$ is said to be \emph{efficient} iff there is no agreement in the feasible set $\mathcal{G}$ that makes every player strictly better off. It is said to be \emph{strongly efficient} or \emph{Pareto optimal} iff there is no other agreement that makes every player at least as well off and at least one player strictly better off. We refer to the set of all efficient agreements as the \emph{efficient frontier} of $\mathcal{G}$. In addition, we refer to the efficient frontier of the individual rational feasible set $\mathcal{G} \cap \{\mathbf{g}|\mathbf{g}\geq \mathbf{g}^0\}$ as the \emph{individual rational efficient frontier}. Given that $\mathcal{G}$ is closed and convex, the regularity conditions in (\ref{eqn:regular1}) and (\ref{eqn:regular2}) hold iff the individual rational efficient frontier is strictly monotone, i.e., it contains no horizonal or vertical line segments. An example illustrating the concepts defined above is shown in Fig. \ref{fig:frontier}. The bargaining problem described in Fig. \ref{fig:frontier} is regular.

We next describe two different bargaining approaches to solving the bargaining problem: NBS and AOBG.
\vspace{0.2cm}
\subsubsection{Nash Bargaining Solution}
This bargaining problem is approached axiomatically by Nash \cite{references:Myerson91}. In this approach, $\mathbf{g}^* = \mathbf{\Phi}(\mathcal{G},\mathbf{g}^0)$ is said to be an NBS in $\mathcal{G}$ for $\mathbf{g}^0$, if the following axioms are satisfied.
\begin{enumerate}
\item Individual Rationality: $\Phi_i(\mathcal{G},\textbf{g}^0) \geq g^0_i, \forall i$
\item Feasibility: $\mathbf{\Phi}(\mathcal{G},\mathbf{g}^0)\in \mathcal{G}$
\item Pareto Optimality: $\mathbf{\Phi}(\mathcal{G},\mathbf{g}^0)$ is Pareto optimal.
\item Independence of Irrelevant Alternatives: For any closed convex set $\mathcal{G}'$, if $\mathcal{G}' \subseteq \mathcal{G}$ and $\mathbf{\Phi}(\mathcal{G},\mathbf{g}^0) \in \mathcal{G}'$,  then $\mathbf{\Phi}(\mathcal{G}',\mathbf{g}^0) = \mathbf{\Phi}(\mathcal{G},\mathbf{g}^0)$.
\item Scale Invariance: For any numbers $\lambda_1, \lambda_2,\gamma_1$ and $\gamma_2$, such that $\lambda_1 > 0$ and $\lambda_2>0$, if $\mathcal{G}' = \{(\lambda_1 g_1+ \gamma_1,\lambda_2 g_2 + \gamma_2)|(g_1,g_2)\in \mathcal{G}\}$ and $\mathbf{\omega} =(\lambda_1 g^0_1+ \gamma_1,\lambda_2 g^0_2 + \gamma_2) $, then $\mathbf{\Phi}(\mathcal{G}',\mathbf{\omega}) = (\lambda_1 \Phi_1(\mathcal{G},\mathbf{g}^0)+ \gamma_1,\lambda_2 \Phi_2(\mathcal{G},\mathbf{g}^0) + \gamma_2)$.
\item Symmetry: If $g^0_1 = g^0_2$, and $\{(g_2,g_1)|(g_1,g_2)\in \mathcal{G}\} = \mathcal{G}$, then $\Phi_1(\mathcal{G},\mathbf{g}^0) = \Phi_2(\mathcal{G},\mathbf{g}^0)$.
\end{enumerate}
Axioms (4)-(6) are also called {\em axioms of fairness}.

{\theorem\cite{references:Myerson91} There is a unique solution $\mathbf{g}^* = \mathbf{\Phi} (\mathcal{G}, \mathbf{g}^0)$ that satisfies all of the above six axioms. This solution is given by,
\begin{equation}
\mathbf{\Phi} (\mathcal{G}, \mathbf{g}^0) = \arg \max _{\mathbf{g} \in \mathcal{G}, \mathbf{g} \geq \mathbf{g}^0}\prod _{i =1}^2 (g_i - g^0_i)\label{eqn:nashproduct}
\end{equation}
}

The NBS selects the unique allocation that maximizes the Nash product in (\ref{eqn:nashproduct}) over all feasible individual rational allocations in $\mathcal{G}\cap \{\mathbf{g}|\mathbf{g}\geq \mathbf{g}^0\}$. Note that for any essential bargaining problem, the Nash point should always satisfy $g^*_i > g^0_i, \forall i$.

\subsubsection{The Bargaining Game of Alternating Offers}
In the cooperative approach to the bargaining problem $(\mathcal{G}, \mathbf{g}^0)$, the NBS is the solution that satisfies a list of properties such as Pareto optimality and fairness. However, using this approach, most information concerning the bargaining environment and procedure is abstracted away, and each user' cost of delay in bargaining is not taken into account. A dynamic noncooperative model of bargaining called the \emph{alternating-offer bargaining game}, on the other hand, provides a detailed description of the bargaining process. In the AOBG, two users take turns in making proposals of payoff allocation in $\mathcal{G}$ until one is accepted or negotiation breaks down.

An important issue regarding modeling of the AOBG is about players' cost of delay in bargaining, as they are directly related to users' motives to settle in an agreement rather than insist indefinitely on incompatible demands. Two common motivations are their sensitivity to time of delay in bargaining and their fear for the risk of breakdown of negotiation\cite{references:Binmore86}. In the bargaining game we consider in this paper, we derive users' cost of delay in bargaining from an exogenous risk of breakdown; i.e., after each round, the bargaining process may terminate in disagreement permanently with an exogenous positive probability if the proposal made in that round gets rejected. In a wireless network, this probability could correspond to the event that other users present in the environment intervene and snatch the opportunity of negotiation on transmission strategies between a pair of users. For example, consider an uncoordinated environment when multiple users operate over a common channel. By default each user's receiver only decodes the intended message from its transmitter and treats the other users' signals as noise. However, groups of users are allowed to coordinate their transmission strategies to improve their respective rates. In the case of a two-user group, if one user's proposal gets rejected by the other user in any bargaining round, it is reasonable to assume that it may terminate the bargaining process and turn to a third user for negotiation. The succeeding analysis for the AOBG with risk of breakdown is based on an extensive game with perfect information and chance moves from game theory \cite{references:Martin}. For completeness, a review of the related concepts from game theory is given in Appendix A.

Consider a regular bargaining problem $(\mathcal{G},\mathbf{g}^0)$ and the two players involved play a dynamic noncooperative game to determine an outcome. Let $p_1$ and $p_2$ be the probabilities of breakdown that satisfy $0<p_1<1$ and $0<p_2<1$. These probabilities of breakdown reflect the users' cost of delay in bargaining and are assumed to be known by both users. The bargaining procedure of this game is as follows. Player 1 and player 2 alternate making an offer in every odd-numbered round and every even-numbered round respectively. An offer made in each round can be any agreement in the feasible set $\mathcal{G}$. Within each round, after the player whose turn it is to offer announces the proposal, the other player can either accept or reject. In any odd-numbered round, if player 2 rejects the offer made by player 1, there is a probability $p_1$ that the bargaining will end in the disagreement $\mathbf{g}^0$. Similarly, in any even-numbered round, if player 1 rejects the offer made by player 2, there is a probability $p_2$ that the bargaining will end in the disagreement $\mathbf{g}^0$. This process begins from round 1 and continues until some offer is accepted or the game ends in disagreement. When an offer is accepted, an agreement is applied and thus the users get the payoffs specified in the accepted offer. Note in the game described above, the two players only get payoffs at a single round in this game, which is the round at which the bargaining ends in either agreement or disagreement. 

A formal description of the above process in the context of an extensive game with perfect information and chance moves introduced in Appendix A is as follows. The player set is $N = \{1,2\}$. Let $T = \{1,2,3,...\}$ denote the index set of bargaining rounds. There is no limit on the number of bargaining rounds. We denote the offer made at round $t$ as $o(t)$. The set of histories $H$ is the set of all sequences of one of the following types:
\begin{enumerate}[I]
\item $\emptyset$, or $(o(1),Re,Cn,o(2),Re,Cn,...,o(t),Re,Cn)$
\item $(o(1),Re,Cn,o(2),Re,Cn,...,o(t))$
\item $(o(1),Re,Cn,o(2),Re,Cn,...,o(t),Ac)$
\item $(o(1),Re,Cn,o(2),Re,Cn,...,o(t),Re)$
\item $(o(1),Re,Cn,o(2),Re,Cn,...,o(t),Re,Br)$
\item $(o(1),Re,Cn,o(2),Re,Cn,...)$
\end{enumerate}
where $t\in T$, $o(t) \in \mathcal{G}$ for all $t$, $Ac$ means ``accept'', $Re$ means ``reject'', $Cn$ means bargaining continues and $Br$ means ``breakdown''.
Histories of Type III, type V and type VI are terminal and those of type VI are infinite. Given a nonterminal history $h$, the player whose turn it is to take an action chooses an agreement in $\mathcal{G}$ as a proposal after a history of type I, chooses a member of $\{Ac,Re\}$ after a history of type II and chooses a member of $\{Cn,Br\}$ after a history of type IV. The player function specifying which player takes an action after a history $h$ is given by: $P(h) = 1$ if $h$ is of either type I or type II and $t$ is even or if $h$ is empty; $P(h) = 2$ if $h$ is of either type I or type II and $t$ is odd; $P(h) = c$ (it is ``chance'''s turn to move) if $h$ is of type IV. For each $h\in H$ with $P(h) = c$, the probability measure $f_c(\cdot|h)$ is given by: $f_c(Br|h) = p_1$ and $f_c(Cn|h) = 1- p_1$ if $h$ is of type IV and $t$ is odd; $f_c(Br|h) = p_2$ and $f_c(Cn|h) = 1- p_2$ if $h$ is of type IV and $t$ is even. Player $i$'s strategy $s_i$ in the game specifies its action to take at any stage of the game when it is its turn to move. When chance moves are present, we need to specify the players' preferences $(\succeq_i)$ over the set of lotteries\footnote{Recall that, from Appendix A, when there are chance moves, the outcome of a strategy profile $s = (s_i)_{i\in N}$ is a probability distribution (or a lottery) over a set of terminal histories instead of a single terminal history.} over terminal histories. We assume these preferences depend only on the final agreements\footnote{If the terminal history $h$ is of type III, the agreement is the last offer $o(t)$ in $h$; if $h$ is of type V instead, the agreement is the disagreement point $\mathbf{g}^0$. Also note that terminal histories of type VI do not occur with positive probability.} reached in the terminal histories of lotteries and not on the path of rejected agreements that preceded them. Moreover, player $i$'s preference relation $\succeq_i$ over the set of all feasible agreements $\mathcal{G}$ can be represented by its payoff $g_i$ where $\mathbf{g} \in \mathcal{G}$.
{\theorem For any regular two-player bargaining problem $(\mathcal{G},\mathbf{g}^0)$, the corresponding AOBG described above has a unique subgame perfect equilibrium (SPE). Let $(\bar{\mathbf{g}},\tilde{\mathbf{g}})$ be the unique pair of efficient agreements in $\mathcal{G}$ which satisfy
\begin{equation}
\tilde{g}_1 = (1-p_2)(\bar{g}_1-g^0_1) + g^0_1 \label{eqn:aobg1}
\end{equation}
\begin{equation}
\bar{g}_2 = (1-p_1)(\tilde{g}_2-g^0_2) + g^0_2\label{eqn:aobg2}
\end{equation}
Let $o_i(t)$ denote user $i$'s payoff in the offer made in round $t$. In the subgame perfect equilibrium, the strategy of player 1 is given by
\begin{equation}
s_1(h) =
\begin{cases}
\bar{\mathbf{g}}\quad & \text{if }h \text{ is of type I and } t \text{ is even}\\
Ac \quad & \text{if }h \text{ is of type II, } t \text{ is even, and } o_1(t)\geq \tilde{g}_1\\
Re \quad & \text{if }h \text{ is of type II, } t \text{ is even, and } o_1(t)< \tilde{g}_1\\
\end{cases}
\end{equation}
and that of player 2 is given by
\begin{equation}
s_2(h) =
\begin{cases}
\tilde{\mathbf{g}}\quad & \text{if }h \text{ is of type I and } t \text{ is odd}\\
Ac \quad & \text{if }h \text{ is of type II, } t \text{ is odd, and } o_2(t)\geq \bar{g}_2\\
Re \quad & \text{if }h \text{ is of type II, } t \text{ is even, and } o_2(t)< \bar{g}_2\\
\end{cases}
\end{equation}
That is, player 1 always proposes an offer $\bar{\mathbf{g}}$ and accepts any offer $\mathbf{g}$ with $g_1 \geq \tilde{g}_1$; user 2 always proposes an offer $\tilde{\mathbf{g}}$ and accepts any offer $\mathbf{g}$ with $g_2 \geq \bar{g}_2$. Using these strategies, the outcome of the game is simply a single terminal history $(\bar{\mathbf{g}},Ac)$. Therefore, in equilibrium, the game will end in an agreement on $\bar{\mathbf{g}}$ at round 1.
}
\begin{proof}
The proof of this theorem is similar to that of Theorem 8.3 in \cite{references:Myerson91} with the disagreement outcome fixed to $\mathbf{g}^0$ after the breakdown in any round. Regularity of the bargaining problem is essential for the proof of the uniqueness of the subgame perfect equilibrium.
\end{proof}

In \cite{references:Binmore86}, it is found that as $p_1$ and $p_2$ approach to zero, the equilibrium outcome of the AOBG converges to the NBS. In other words, if there are no external forces to terminate the bargaining process, the equilibrium outcome of the dynamic game approaches the NBS. More discussion will be given on how the probabilities of breakdown $p_1$ and $p_2$ affect the equilibrium outcome of the bargaining game in the later sections.

For convenience, Table \ref{table:notations} summarizes various notations used in this subsection.
\begin{table}\footnotesize
\centering
\begin{tabular}{ | l | l |}
    \hline
    \textbf{Notations} & \textbf{Meanings} \\ \hline
    $\mathcal{G}$ & feasible set  \\ \hline
    $g_i$ & user $i$'s payoff in agreement $\mathbf{g}$ \\ \hline
    $\mathbf{g}^0$ & disagreement point\\ \hline
    $\mathbf{g}^*$, $\mathbf{\Phi}(\mathcal{G}, \mathbf{g}^0)$ & NBS of $(\mathcal{G}, \mathbf{g}^0)$\\ \hline
    $p_i$ & probability of breakdown when user $i$'s offer is rejected \\ \hline
    $o(t)$ & offer made at round $t$ \\ \hline
    $H$ & set of histories (defined in Appendix A) \\ \hline
    $h$ & a history (defined in Appendix A)\\ \hline
    $P(h)$ & player function (defined in Appendix A)\\ \hline
    $f_c(\cdot|h)$ & probability measure (defined in Appendix A) \\ \hline
    $s_i(h)$ & player $i$'s strategy (defined in Appendix A)\\ \hline
    $Ac$ & accept \\ \hline
    $Re$ & reject \\ \hline
    $Br$ & breakdown \\ \hline
    $\bar{\mathbf{g}}$ & offer of user $1$ in subgame perfect equilibrium \\ \hline
    $\tilde{\mathbf{g}}$ & offer of user $2$ in subgame perfect equilibrium \\
    \hline
\end{tabular}
\caption{Notations used in Section II-C.}
\label{table:notations}
\end{table}
\section{Bargaining over the Two-User Gaussian MAC}
Before we move to the Gaussian IC, we first illustrate the bargaining framework for a Gaussian MAC in which two users send information to one common receiver. Cooperative bargaining using the NBS has been discussed before for the MAC in \cite{references:Mathur_Sankar_Mandayam06}. In this section, we reconsider the bargaining problem in the two-user case and provide a closed-form solution for the NBS. Besides, we also study the bargaining outcome when a noncooperative bargaining approach is used. The results here also form the foundation for the solution of the strong IC, which will be studied later. The channel is
\begin{equation}
Y_t = X_{1,t} + X_{2,t} + Z_t
\end{equation}
where $X_{i,t}$ is the input of user $i$, $Y_{t}$ is the output and $Z_t$ is i.i.d. Gaussian noise with zero mean and unit variance at time $t= 1,2,...,n$. Each user has an individual average input power constraint $P_i$ given by (\ref{eqn:powerconst}). The capacity region $\mathcal{C}_0$ is the set of all rate pairs $(R_1, R_2)$ such that
\begin{eqnarray}
R_i \leq C(P_i), \: i \in \{1,2\}\\
R_1 + R_2 \leq C(P_1 + P_2) = \phi_0
\end{eqnarray}
If the two users fully cooperate in codebook and rate selection, any point in $\mathcal{C}_0$ is achievable. When there is no coordination between users, in the worst case, one user's signal can be treated as noise in the decoding of the other user, leading to rate $R_i^0 = C(\frac{P_i}{1+P_{3-i}})$  for user $i$. In \cite{references:Gajic08}, $R_i^0$ is also called user $i$'s ``safe rate''. If the two users are selfish but willing to coordinate for mutual benefits, they may bargain over $\mathcal{C}_0$ to obtain a preferred operating point with $\mathbf{R}^0$ serving as a disagreement point. In the following, we focus on how to find the solution to the bargaining problem $(\mathcal{C}_0,\mathbf{R}^0)$ using both the NBS approach and the AOBG approach respectively.

\subsection{The NBS Approach}
It can be easily observed that the feasible set $\mathcal{C}_0$ in the MAC case is bounded by only three linear constraints on $R_1$ and $R_2$. Before we move to determine the NBS in the MAC case, we first solve the NBS to the bargaining problem with a more general feasible set $\mathcal{G}$ and a particular disagreement point $\mathbf{g}^0$, the results of which will also be useful for the IC case in Section IV and Section V. We assume the feasible set $\mathcal{G}$ has the following general form:
\begin{equation}
\mathcal{G} = \{\mathbf{g}\in \mathbb{R}^2 |\mathbf{g} \geq \mathbf{0},\;\mathbf{g}\leq \mathbf{g}^1\; \text{and}\; \mathbf{A}\mathbf{g}\leq \mathbf{B}\}
\end{equation}
where $\mathbf{g}^1$ is a $2\times 1$ vector that contains the maximum possible payoff for each user, the $J\times 2$ matrix $\mathbf{A} = (A_{ji})$ and the $J\times 1$ vector $\mathbf{B}$ are related to the $J$ linear constraints.

{\proposition Assuming that $\mathbf{g}^0<\mathbf{g}^1\:\text{and} \:\mathbf{A}\mathbf{g}^0<\mathbf{B}$, there exists a unique NBS $\mathbf{g}^*$ for the bargaining problem $(\mathcal{G},\mathbf{g}^0)$, which is given by

\begin{equation}
g_i^* =  \min \left\{ g_i^1; g_i^0 +\frac{1}{\sum_{j = 1}^J \mu_j A_{ji}}\right\},\quad i\in\{1,2\}
\end{equation}
where the Lagrange multipliers $\mu_j\geq 0$ ($j\in\{1,...,J\}$) are found by solving $(\mathbf{Ag}^*-\mathbf{B})_j \mu_j = 0$ and $\mathbf{A}\mathbf{g}^* \leq \mathbf{B}$.
}
\begin{proof}
Maximizing the Nash product in (\ref{eqn:nashproduct}) is equivalent to maximizing its logarithm.
Define $m(\mathbf{g}) = \ln(g_1 - g_1^0) + \ln(g_2 - g_2^0)$, then $m(\cdot): \mathcal{G}\cap \{\mathbf{g}|\mathbf{g} \geq \mathbf{g}_0\} \rightarrow \mathbb{R}^+$ is a strictly concave function of $\mathbf{g}$. Also note that the constraints $(Ag)_j \leq B_j,\; j \in \{1,2,...,J\}$ are linear in $g_1$ and $g_2$. So the first order Karush-Kuhn-Tucker conditions are necessary and sufficient for optimality \cite{references:Bertsekas}.
Let $L(\mathbf{g},\mathbf{\lambda},\mathbf{\nu},\mathbf{\mu})$ denote the Lagrangian function and $\lambda_i \geq 0,\: i = 1,2$, $\nu_i \geq 0,\: i = 1,2$ and $\mu_j \geq 0,\: j = 1,2,...,J$ denote the Lagrange multipliers associated with the constraints, then we have
\begin{equation}
\begin{array}{l}
L(\mathbf{g},\mathbf{\lambda},\mathbf{\nu},\mathbf{\mu}) = m(\mathbf{g}) + \sum_{i = 1}^2 \lambda_i(g_i - g_i^0)
 +\sum_{i = 1}^2 \nu_i(g_i^1-g_i) + \sum_{j=1}^J \mu_j (B_j - (Ag)_j)
\end{array}
\end{equation}
The first-order necessary and sufficient conditions yield
\begin{equation}
1 + \left(\lambda_i - \nu_i - \sum_{j=1}^J \mu_j A_{ji}\right)(g_i^* - g_i^0) = 0; \: i = 1,2
\end{equation}
and
\begin{eqnarray}
(g_i^* - g_i^0)\lambda_i = 0; \quad \lambda_i \geq 0; \: i = 1,2\\
(g_i^1-g^*_i)\nu_i = 0; \quad \nu_i \geq 0; \: i = 1,2\\
((Ag^*)_j-B_j)\mu_j = 0; \quad \mu_l \geq 0;\: j = 1,2,...,J
\end{eqnarray}
Since $g_i^* > g_i^0$ must hold, we have $\lambda_i = 0$ for $i = 1,2$. In addition, if $g_i^*<g_i^1$, then $\nu_i = 0$; otherwise $g_i^* = g_i^1$. Thus, the results in Proposition 1 follow.
\end{proof}
In the MAC case, we have $\mathcal{G} = \mathcal{C}_0$, $\mathbf{g}^0 = (C(\frac{P_1}{1+P_2})\;C(\frac{P_2}{1+P_1}))^t$, $\mathbf{g}^1 = (C(P_1)\;C(P_2))^t$, $A = (1\; 1)$ and $B = \phi_0$ in Proposition 1. Note the conditions $\mathbf{g}^0<\mathbf{g}^1\:\text{and} \:\mathbf{A}\mathbf{g}^0<\mathbf{B}$ always hold; i.e., both users operating over the MAC always have incentives to cooperate. Since the only linear constraint is always active (i.e., $\mu_1>0$), the optimization problem can be solved fully and has a closed-form solution as summarized in the following proposition.
{\proposition There exists a unique NBS for the two-user Gaussian MAC bargaining problem $(\mathcal{C}_0, \mathbf{R}^0)$, given by

\begin{equation}
\mathbf{R}^* = (R_1^0+\frac{1}{\mu_1}\;\: R_2^0+\frac{1}{\mu_1})^t
\end{equation}
 where $\mu_1 = \frac{2}{\phi_0-R_1^0-R_2^0}$.
}

\subsection{The AOBG Approach}
In this subsection, we apply the AOBG framework to the case of two-user MAC and analyze the negotiation results.

For the two-user MAC bargaining problem $(\mathcal{C}_0,\mathbf{R}^0)$, the individual rational efficient frontier is strictly monotone and thus the regularity conditions in Section II always hold. Hence, using Theorem 2, we have the following proposition.
{\proposition For the two-user MAC bargaining problem $(\mathcal{C}_0,\mathbf{R}^0)$, the unique pair of agreements $(\bar{\mathbf{R}}, \tilde{\mathbf{R}})$ in the subgame perfect equilibrium of the AOBG is given by
\begin{equation}
(\bar{R}_1\;\bar{R}_2\;\tilde{R}_1\;\tilde{R}_2)^t = M^{-1}(-p_2R_1^0\:\:p_1 R_2^0\:\: \phi_0\:\: \phi_0)^t\label{eqn:spemac}
\end{equation}
where
\begin{equation}
M = \begin{pmatrix}
  1-p_2 & 0 & -1 & 0 \\
  0 & 1 & 0 & -(1-p_1) \\
  1 & 1  & 0 & 0  \\
  0 & 0 & 1 & 1
 \end{pmatrix}
\end{equation}
In equilibrium, the game will end in an agreement on $\mathbf{\bar{R}}$ at round 1.
}
\begin{proof}
From (\ref{eqn:aobg1}) and (\ref{eqn:aobg2}) in Theorem 2, it follows that the unique pair of agreements $(\bar{\mathbf{R}}, \tilde{\mathbf{R}})$ in the subgame perfect equilibrium must satisfy
\begin{equation}
\tilde{R}_1 = (1-p_2)(\bar{R}_1-R^0_1) + R^0_1 \label{eqn:aobgmac1}
\end{equation}
\begin{equation}
\bar{R}_2 = (1-p_1)(\tilde{R}_2-R^0_2) + R^0_2\label{eqn:aobgmac2}
\end{equation}
In addition, since $\bar{\mathbf{R}}$ and $\tilde{\mathbf{R}}$ need to be efficient agreements, we have
\begin{equation}
\bar{R}_1 + \bar{R}_2 = \phi_0 \label{eqn:aobgmac3}
\end{equation}
\begin{equation}
\tilde{R}_1 + \tilde{R}_2 = \phi_0 \label{eqn:aobgmac4}
\end{equation}
Solving (\ref{eqn:aobgmac1}), (\ref{eqn:aobgmac2}), (\ref{eqn:aobgmac3}) and (\ref{eqn:aobgmac4}), we obtain the unique pair of agreements $(\bar{\mathbf{R}}, \tilde{\mathbf{R}})$ as in the proposition.
\end{proof}

Clearly, if user 2 makes an offer during the first round instead, the equilibrium outcome would be $\tilde{\mathbf{R}}$. It is not hard to see from (\ref{eqn:aobgmac1}), (\ref{eqn:aobgmac2}) that if $p_1 = p_2 = 0$, then we have $\tilde{\mathbf{R}} = \bar{\mathbf{R}}$.

In Fig. \ref{fig:macnbs}, the capacity region, the disagreement point and the NBS obtained using Proposition 2 are illustrated for $\text{SNR}_1 = 20$dB and $\text{SNR}_2 = 15$dB. Recall that the mixed strategy NE in \cite{references:Gajic08} has an average performance equal to the safe rates in $\mathbf{R}^0$. The NBS point which is the unique fair Pareto-optimal point in $\mathcal{C}_0$ is component-wise superior. This shows that bargaining can improve the rates for both selfish users in a MAC. Also included are the unique pairs of agreements $(\bar{\mathbf{R}},\tilde{\mathbf{R}})$ in Proposition 3 for two different choices of $p_1$ and $p_2$. Recall that offer of user 1 in subgame perfect equilibrium $\bar{\mathbf{R}}$ corresponds to the equilibrium outcome of the AOBG since we assume user 1 makes an offer first. If user 2 is the first mover instead, offer of user 2 in subgame perfect equilibrium $\tilde{\mathbf{R}}$ becomes the equilibrium outcome of the game. For a fixed pair of $p_1$ and $p_2$, each user's rate in the equilibrium outcome is higher when it is the first mover than when it is not. Such a phenomenon is referred to as ``first mover advantage'' in \cite{references:Martin}. Finally, as shown in the figure, when $p_1$ and $p_2$ become smaller, both $\tilde{\mathbf{R}}$ and $\bar{\mathbf{R}}$ are closer to the Nash solution.

\section{Two-User Gaussian IC}
For a general Gaussian IC, the capacity region is not known. While the full H-K rate region \cite{references:Han81} gives the largest known achievable rate region, as discussed in Section II-B, taking into account all possible power splits and different time-sharing strategies makes it computationally infeasible. For tractability, we consider a simple H-K type scheme with fixed power split and no time-sharing. For the strong interference case, we set $\alpha = \beta = 0$, which is known to be optimal\cite{refereces:Sato81}. For the weak and mixed interference cases, we choose the near-optimal power splits of \cite{references:Etkin08}. For weak interference $a<1$ and $b<1$, we set $\alpha = \min(1/(bP_1),1)$ and $\beta = \min(1/(aP_2),1)$; for mixed interference $a<1$ and $b\geq 1$, we set $\alpha = 0$ and $\beta = \min(1/(aP_2),1)$. In the uncoordinated case, each receiver treats the interfering signal as noise, leading to rates in disagreement point $\mathbf{R}^0= (C(\frac{P_1}{1+aP_2})\;C(\frac{P_2}{1+bP_1}))^t$. 

The simple H-K scheme discussed above requires each user to split its rate for the benefit of both users. However, it is not always true that each user will be able to improve its rate over the disagreement point $\mathbf{R}^0$ as a result of the employed simple H-K scheme and the resulting bargaining problem will be essential as defined in Section II-C. In order to ensure that both selfish users will have motives to employ H-K coding, a pre-bargaining phase is added before the actual bargaining phase. We refer to this pre-bargaining phase as phase 1 and the bargaining phase that follows as phase 2.

In phase 1, users check whether the simple H-K scheme improves individual rates for both over those in disagreement $\mathbf{R}^0$. If there is no improvement for at least one user, then that user does not have the incentive to cooperate and negotiation breaks down. In such a scenario, users operate at the disagreement point $\mathbf{R}^0$. Otherwise, they reach an agreement on the use of the simple H-K scheme with the chosen power split and proceed to phase 2. In phase 2, the users bargain for a rate pair to operate at over the achievable rate region of the H-K scheme they agreed on earlier. The second phase can then be formulated as a two-user bargaining problem with the feasibility set $\mathcal{F}$ defined in Section II-B and disagreement point $\mathbf{R}^0$. Once a particular rate pair is determined as the solution of the second phase bargaining problem, related codebook information is shared between the users so that one user's receiver can decode the other user's common message as required by the adopted H-K scheme. If negotiation breaks down, in phase 2, the receivers are not provided with the interfering user's codebook.

\subsection{Phase 1: the Pre-bargaining Phase}

In this subsection, we discuss the pre-bargaining phase and study conditions under which both users have incentives to engage in the use of the simple H-K scheme discussed above.

{\proposition For the two-user Gaussian IC, the pre-bargaining phase is successful and both users have incentives to employ an H-K scheme provided one of the following conditions hold. The conditions also list the H-K scheme employed by the users.
\begin{itemize}
\item Strong interference ($a \geq 1$ and $b \geq 1$): Users always employ HK(0,0);
\item Weak interference ($a <1$ and $b<1$): Users employ HK($1/(bP_1)$,$1/(aP_2)$) iff $aP_2>1$ and $bP_1 > 1$ and $\mathcal{F}\cap \{\mathbf{R}>\mathbf{R}^0\}$ is nonempty when $\alpha = 1/(bP_1)$ and $\beta = 1/(aP_2)$;
\item Mixed interference ($a < 1$ and $b \geq 1$): Users employ HK($0$,$1/(aP_2)$) iff $aP_2>1$ and $\mathcal{F}\cap \{\mathbf{R}>\mathbf{R}^0\}$ is nonempty when $\alpha = 0$ and $\beta = 1/(aP_2)$.
\end{itemize}
\label{thm:incentive}
}
\begin{proof}
See Appendix B.
\end{proof}

Note that in the weak and mixed interference cases, when both $\text{SNR}$'s are high, the conditions $aP_2 > 1$ and $bP_1 > 1$ are satisfied for most channel gains and it only remains to check whether $\mathcal{F}\cap \{\mathbf{R}>\mathbf{R}^0\}$ is nonempty. This implies that in the interference limited regimes, it is very likely that both users would have incentives to cooperate.

\subsection{Phase 2: the Bargaining Phase}
\subsubsection{Nash Bargaining Solution over IC}
After the users agree on an H-K scheme, in phase 2, if bargaining is cooperative, the NBS over the corresponding rate region $\mathcal{F}$ is employed as the operating point. Since the pre-bargaining in phase 1 is successful, we concentrate on the case when $\mathbf{R}^0<\mathbf{R}^1\:\text{and} \:\mathbf{A}_0\mathbf{R}^0<\mathbf{B}_0$ for the chosen HK($\alpha$, $\beta$) scheme and $\mathcal{F}\cap \{\mathbf{R}>\mathbf{R}^0\}$ is nonempty.
Applying Proposition 1 with the feasible set $\mathcal{F}$ and the disagreement point $\mathbf{R}^0$, we have the following result.
{\proposition Provided the pre-bargaining phase is successful, there exists a unique NBS for the bargaining problem $(\mathcal{F}, \mathbf{R}^0)$ in phase 2, which is characterized in Proposition 1 with $\mathcal{G} = \mathcal{F}$, $\mathbf{g}^0 = \mathbf{R}^0 = (C(\frac{P_1}{1+aP_2})\;C(\frac{P_2}{1+bP_1}))^t$, $\mathbf{g}^1 = (C(P_1)\;C(P_2))^t$, $\mathbf{A} = \mathbf{A}_0$ and $\mathbf{B} = \mathbf{B}_0$.
}

We will elaborate on the NBS in Section IV-C.
\subsubsection{Alternating-Offer Bargaining Games over IC}

If bargaining is noncooperative in phase 2, analysis for the AOBG over the IC is similar to that over the MAC in the Section III; however, unlike in the MAC case, the associated bargaining problem over the IC is not always regular. If it is non-regular, the AOBG may have more than one subgame perfect equilibria resulting in distinct bargaining outcomes, which puts any of the subgame perfect equilibria and the corresponding outcome in doubt \cite{references:Myerson91}. Hence the non-regular case is not treated here. In the following, we discuss the regularity of the associated bargaining problem in different interference regimes and characterize the unique subgame perfect equilibrium of the AOBG when the bargaining problem is regular.

{\proposition Provided the pre-bargaining phase is successful, in phase 2, the two-user Gaussian IC bargaining problem $(\mathcal{F}, \mathbf{R}^0)$ is regular iff one of the following conditions hold:
\begin{itemize}
\item Strong interference: $a = b = 1$;
\item Weak interference: $R_1^0 \geq (\phi_5-2\phi_2)^+$ and $R_2^0 \geq (\phi_4-2\phi_1)^+$;

\item Mixed interference: $R_1^0 \geq (\min(\phi_5-2\phi_2,\phi_3-\phi_2))^+$ and $R_2^0 \geq (\min(\phi_4-2\phi_1,\phi_3-\phi_1))^+$;
\end{itemize}
where $\phi_i, i=1,...,5$ are defined in (\ref{eqn:reg1})-(\ref{eqn:reg5}).
\label{thm:regularity}
}
\begin{proof}
See Appendix C.
\end{proof}

Fig. \ref{fig:regular_range} shows the set of cross-link power gains $(a,b)$ for which the associated bargaining problem is regular. Note the conditions for regularity not only include those in Proposition \ref{thm:regularity} but also those in Proposition \ref{thm:incentive} as well since we assume the pre-bargaining phase has been successful. In Fig. \ref{fig:regular_range1}, we have $\text{SNR}_1 = \text{SNR}_2 = 20$dB. We observe that $(\mathcal{F}, \mathbf{R}^0)$ is regular for a large range of power gains in the weak interference regime. In Fig. \ref{fig:regular_range2}, we set $\text{SNR}_1 = 20$dB and $\text{SNR}_2 = 30$dB, and observe that, in addition to part of the weak interference regime,  $(\mathcal{F}, \mathbf{R}^0)$ is also regular for a range of power gains in the mixed interference regime. Besides, in both scenarios, the bargaining problem is regular for the special case of strong interference $a=b=1$. Finally, note that in the noisy interference regime when $a$, $b$, $P_1$ and $P_2$ satisfy $\sqrt{a}(bP_1+1)+\sqrt{b}(aP_2+1)\leq 1$ \cite{references:Shang09}, since treating interference as noise is optimal, users never employ the H-K scheme and the pre-bargaining phase always fails.

When pre-bargaining in phase 1 is successful and the Gaussian IC bargaining problem $(\mathcal{F},\mathbf{R}^0)$ is regular, using Theorem 2, we have the following result.
{\proposition  For any regular bargaining problem $(\mathcal{F},\mathbf{R}^0)$ over the two-user Gaussian IC, the unique pair of agreements $(\bar{\mathbf{R}}, \tilde{\mathbf{R}})$ in the subgame perfect equilibrium of the AOBG both lie on the individual rational efficient frontier of $\mathcal{F}$ and satisfy (\ref{eqn:aobg1}) and (\ref{eqn:aobg2}) with $\mathbf{R}^0 = (C(\frac{P_1}{1+aP_2})\;C(\frac{P_2}{1+bP_1}))^t$.
\label{thm:spe_ic}
}

In the strong interference case $a = b = 1$, the unique pair of agreements $(\bar{\mathbf{R}}, \tilde{\mathbf{R}})$ in the subgame perfect equilibrium can be obtained using (\ref{eqn:spemac}) in Proposition 3 with $\phi_0$ replaced by $\phi_6$. For the weak and mixed interference cases, since the shape of the H-K rate region and the relative location of the disagreement point vary as parameters $a$, $b$, $P_1$ and $P_2$ change, it is difficult to obtain a general expression for $(\bar{\mathbf{R}}, \tilde{\mathbf{R}})$. However, when all the parameters are given and the corresponding power split parameters $\alpha$ and $\beta$ are fixed, the H-K rate region and the disagreement point $\mathbf{R}^0$ can be determined accordingly. Since $(\bar{\mathbf{R}}, \tilde{\mathbf{R}})$ both lie on the individual rational efficient frontier of $\mathcal{F}$ which is piecewise linear, we can compute $(\bar{\mathbf{R}}, \tilde{\mathbf{R}})$ by solving linear equations.


\subsection{Illustration of Results}
The achievable rate region of the H-K scheme with the optimal or near-optimal power split discussed earlier and the corresponding NBS (we refer to it as H-K NBS) together with disagreement points are plotted for different values of channel parameters in Fig. \ref{fig:hkvstdm}. For comparison, we also include the TDM regions and the corresponding NBS (we refer to it as TDM NBS). When TDM is employed, user $i$ transmits a fraction $\rho_i (0\leq \rho_i\leq 1)$ of the time under the constraint $\rho_1 + \rho_2 \leq 1$. For a given vector $\mathbf{\rho} = (\rho_1\;\rho_2)^t$, the rate obtained by user $i$ is given by $R_i(\mathbf{\rho}) = R_i(\rho_i) = \rho_iC(\frac{P_i}{\rho_i})$. Hence, the TDM rate region is given by $\mathcal{R}_{\text{TDM}} = \{\mathbf{R}|\mathbf{R} = (R_1(\rho_1)\;R_2(\rho_2))^t,\; \rho_1+\rho_2\leq 1\}$ and the TDM NBS is computed as the solution to the bargaining problem $(\mathcal{R}_{\text{TDM}},\mathbf{R}^0)$. The NBS based on TDM was also investigated for a Gaussian interference game in \cite{references:Leshem08} using the unique competitive solution studied there as the disagreement point. Note that for TDM Proposition 5 applies and since the efficient frontier of the TDM rate region is strictly monotone, the associated bargaining problem is regular as long as it is essential.

Since interference limited regimes are more of interest here, in these plots, we assume the signal to noise ratios for both users' direct links are high, i.e, $\text{SNR}_1 = \text{SNR}_2 = 20$dB. In each case, the channel parameters are chosen according to Proposition \ref{thm:incentive} so that the pre-bargaining phase is successful. In Fig. \ref{fig:subfig1}, both interfering links are strong, $\text{HK}(0,0)$ is employed. The H-K NBS strictly dominates the TDM one. Fig. \ref{fig:subfig2} shows an example for mixed interference with $a = 0.1$ and $b = 3$. Since $aP_2 = 10 > 1$, $\text{HK}(0,0.1)$ is employed. In this example, although TDM results in some rate pairs that are outside the H-K rate region, the H-K NBS remains component-wise better than the TDM one. The weak interference case when $a = 0.2$ and $b = 0.5$ is plotted in Fig. \ref{fig:subfig3}. For these parameters, we have $aP_2 = 20>1$ and $bP_1 = 50>1$, therefore $\text{HK}(0.02,0.05)$ is used. The H-K NBS in this case, though still much better than $\mathbf{R}^0$, is slightly worse than the TDM one. This is because the TDM rate region contains the H-K rate region due to the suboptimality of the simple H-K scheme in the weak regime. Finally, recall that while the TDM rate region does not depend on $a$ and $b$, since $\mathbf{R}^0$ does, the TDM NBS depends on $a$ and $b$ as well.

We compute the H-K NBS for different ranges of the channel parameters in Fig. \ref{fig:ratesvsb}. We assume $\text{SNR}_1 = \text{SNR}_2 = 20$dB, $a = 1.5$ and $b$ varies from 0 to 3. The improvement of each user's rate in $\mathbf{R}^*$ over the one in $\mathbf{R}^0$ increases as $b$ grows. When $b< a$, user 1's rate in the NBS is less than user 2's; however, as $b$ grows beyond $a$, user 1's rate in the NBS surpasses user 2's, which is due to the fairness property of the NBS. Alternatively we say a strong interfering link can give user 1 an advantage in bargaining.

In Fig. \ref{fig:icaobg}, the unique pair of agreements $(\bar{\mathbf{R}}, \tilde{\mathbf{R}})$ in the subgame perfect equilibrium of the AOBG is shown for mixed interference with $a = 0.2$, $b = 1.2$, $\text{SNR}_1 = 10$dB and $\text{SNR}_2 = 20$dB for three different choices of the pair of probabilities of breakdown $p_1$ and $p_2$. According to Proposition \ref{thm:incentive}, in phase 1, the two users decide to cooperate using $\text{HK}(0,0.05)$. Furthermore, by Proposition \ref{thm:regularity}, the bargaining problem in phase 2 is regular. As in the MAC case, user 1's offer in subgame perfect equilibrium $\bar{\mathbf{R}}$ corresponds to the equilibrium outcome of the AOBG since we assume user 1 makes an offer first. If user 2 moves first instead, user 2's offer in subgame perfect equilibrium $\tilde{\mathbf{R}}$ would become the equilibrium outcome of the game. We can see that as $p_1$ and $p_2$ change, $\bar{\mathbf{R}}$ and $\tilde{\mathbf{R}}$ move along the individual rational efficient frontier of $\mathcal{F}$. When $p_1 = 0.5$ and $p_2 = 0.5$, user 1's rate in $\bar{\mathbf{R}}$ is greater than that in the NBS; but when $p_1 = 0.1$ and $p_2 = 0.5$, its rate in $\bar{\mathbf{R}}$ is smaller than that in the NBS. As $p_1$ and $p_2$ decrease to $0.1$, both $\bar{\mathbf{R}}$ and $\tilde{\mathbf{R}}$ become closer to the Nash solution. The rate of each user in the perfect equilibrium outcome $\bar{\mathbf{R}}$ as a function of breakdown probability $p_1$ is plotted in Fig. \ref{fig:speratevsp1} when $p_2$ is fixed to 0.5 under the above channel parameters. As $p_1$ gets larger, user 1's rate increases while user 2's decreases. The larger $p_1$ becomes, the more likely that bargaining may permanently terminate in disagreement when user 1's offer is rejected by user 2. This demonstrates that if user 1 fears less about bargaining breakdown, it can be more advantageous in bargaining. It should also be emphasized that due to regularity the equilibrium is unique and agreement is reached in round 1 in equilibrium. In this sense, the bargaining mechanism of AOBG is highly efficient.

Fig. \ref{fig:aobgtdm} illustrates the perfect equilibrium outcomes of the AOBG when the H-K and TDM cooperating schemes are used respectively for an example of mixed interference with $a = 0.2$, $b = 1.2$, $\text{SNR}_1 = 20$dB and $\text{SNR}_2 = 30$dB. By Proposition's \ref{thm:incentive}, \ref{thm:regularity} and Fig. \ref{fig:regular_range2}, incentive conditions in phase 1 are satisfied and the bargaining problem is regular. The equilibrium outcomes of the AOBG in the TDM case are obtained by applying Proposition \ref{thm:spe_ic}. Since the boundary of the TDM rate region is not linear, we compute the unique pair of $(\bar{\mathbf{R}}, \tilde{\mathbf{R}})$ in TDM numerically. The probabilities of breakdowns are set as $p_1 = p_2 = 0.5$. The NBS's in both cases are also plotted for reference. We observe that the individual rational efficient frontiers for the H-K and TDM schemes intersect. Also, while user 2 gets higher rates in all the bargaining outcomes in TDM than in H-K, user 1's rates in H-K are superior to those in TDM. Hence, we can conclude that, depending on the channel parameters and power constraints, the two users may have distinct preferences between the transmission schemes employed.

\section{Bargaining for the Generalized Degree of Freedom}
In the previous section, we have studied the bargaining problem in which the two selfish users over a Gaussian IC bargain for a fair rate pair over the rate region achieved by the simple H-K scheme. However, for fixed channel parameters $a$, $b$ and power constraints $P_1$ and $P_2$, the employed H-K scheme is a suboptimal one as it can only achieve within one bit to the capacity region in the weak and mixed regimes. In this section, we focus our attention on certain high $\text{SNR}$ regimes when the simple H-K scheme becomes asymptotically optimal and employ the g.d.o.f. as a performance measure for each user. As the g.d.o.f. approximates interference-limited performance well at high SNR's for all interference regimes, the results in this section help us understand what the bargaining solution we would get if bargaining was done over the entire capacity region. Before we deal with the bargaining problem, we briefly review the concept of g.d.o.f. first.

Let $\mathcal{C}(\text{SNR}_1,\text{SNR}_2,\text{INR}_1,\text{INR}_2)$ denote the capacity region of a real Gaussian IC with parameters $\text{SNR}_1$, $\text{SNR}_2$, $\text{INR}_1$ and $\text{INR}_2$ defined in Section II, and let
\begin{eqnarray*}
\theta_1 = \frac{\log \text{SNR}_2}{\log \text{SNR}_1}\\
\theta_2 = \frac{\log \text{INR}_1}{\log \text{SNR}_1}\\
\theta_3 = \frac{\log \text{INR}_2}{\log \text{SNR}_1}
\end{eqnarray*}
Note that for the g.d.o.f analysis, $\theta_1$, $\theta_2$ and $\theta_3$ are fixed\footnote{Note that, to guarantee this, the channel parameters $a$ and $b$ need to change with power $P_1$ and $P_2$.}.
In this section, we only focus on the nontrivial cases when $\theta_i > 0$, $i = 1,2,3$.

The generalized degrees of freedom region is defined as \cite{references:Etkin08}
\begin{align}
\mathcal{D}(\theta_1,\theta_2,\theta_3) = \lim_{\substack{\text{SNR}_1,\text{SNR}_2,\text{INR}_1,\text{INR}_2\rightarrow \infty\\\theta_1,\theta_2,\theta_3\: \text{fixed}}}\{(d_1,d_2)|\nonumber\\
(\frac{d_1}{2}\log \text{SNR}_1,\frac{d_2}{2}\log \text{SNR}_2)\in \mathcal{C}(\text{SNR}_1,\text{SNR}_2,\text{INR}_1,\text{INR}_2)\}
\end{align}
The generalized degrees of freedom $d_1$ and $d_2$ reflect to what extent interference affects communications. When the interference is absent, each user can achieve a rate $R_i = 1/2\log \text{SNR}_i$; as a result of interference, the single-user capacity is scaled by a factor $d_i$. The greater $d_i$ is, the less user $i$ is affected by interference. The following theorem from \cite{references:Etkin08} describes the optimal g.d.o.f. region of a two-user Gaussian IC.

{\theorem
In the strong interference regime $\text{INR}_1 \geq \text{SNR}_2$ and $\text{INR}_2 \geq \text{SNR}_1$ ($\theta_2 \geq \theta_1$ and $\theta_3 \geq 1$), the g.d.o.f. region $\mathcal{D}_1$ is given by
\begin{equation}
d_1\leq 1
\label{eqn:gdof1}
\end{equation}
\begin{equation}
d_2\leq 1
\label{eqn:gdof2}
\end{equation}
\begin{equation}
d_1+\theta_1 d_2 \leq \varphi_1 = \min(\max(1,\theta_2),\max(\theta_1,\theta_3))
\end{equation}
and it is achieved by $\text{HK}(0,0)$.

In the weak interference regime $\text{INR}_1 < \text{SNR}_2$ and $\text{INR}_2 < \text{SNR}_1$ ($\theta_2 < \theta_1$ and $\theta_3 < 1$), the g.d.o.f. region $\mathcal{D}_2$ is given by
(\ref{eqn:gdof1}), (\ref{eqn:gdof2}) and
\begin{equation}
\begin{array}{l l}
d_1+\theta_1 d_2 \leq \varphi_2 = & \min(1+(\theta_1-\theta_3)^+,\theta_1+(1-\theta_2)^+,\\
& \max(\theta_2,1-\theta_3)+\max(\theta_3,\theta_1-\theta_2))\label{eqn:weakdof1}
\end{array}
\end{equation}
\begin{equation}
2d_1+\theta_1 d_2\leq \varphi_3 = \max (1,\theta_2) + \max(\theta_3,\theta_1-\theta_2) + 1-\theta_3\label{eqn:weakdof2}
\end{equation}
\begin{equation}
d_1+2\theta_1 d_2\leq \varphi_4 = \max (\theta_1,\theta_3) + \max(\theta_2,1-\theta_3) + \theta_1-\theta_2\label{eqn:weakdof3}
\end{equation}
and it is achieved by $\text{HK}(1/\text{INR}_2,1/\text{INR}_1)$.

In the mixed interference regime $\text{INR}_1 \geq \text{SNR}_2$ and $\text{INR}_2 < \text{SNR}_1$ ($\theta_2 \geq \theta_1$ and $\theta_3 < 1$), the g.d.o.f. region $\mathcal{D}_3$ is given by
(\ref{eqn:gdof1}), (\ref{eqn:gdof2}) and
\begin{equation}
d_1+\theta_1 d_2 \leq \varphi_5 = \min(1+(\theta_1-\theta_3)^+,\max(1,\theta_2))
\end{equation}
\begin{equation}
d_1+2\theta_1 d_2 \leq \varphi_6 = \max(\theta_1,\theta_3) + \max(\theta_2,1-\theta_3)
\end{equation}
and is achieved by $\text{HK}(1/\text{INR}_2,0)$.
}
\vspace{0.4cm}

Each selfish user aims to merely increase its own g.d.o.f.. If the two users do not coordinate, each user treats the other user's signal as noise. In the uncoordinated case, the pair of rates in disagreement are given by
\begin{eqnarray}
 R_1^0 = \frac{1}{2}\log \left(1+\frac{\text{SNR}_1}{1+\text{INR}_1}\right)\\
 R_2^0 = \frac{1}{2}\log \left(1+\frac{\text{SNR}_2}{1+\text{INR}_2}\right)
 \end{eqnarray}
and thus the corresponding disagreement g.d.o.f. pair $\mathbf{d}^0 = (d_1^0\;d_2^0)^t$ can be obtained as
\begin{equation}
d_1^0 = \lim_{\substack{\text{SNR}_1,\text{SNR}_2,\text{INR}_1,\text{INR}_2\rightarrow \infty\\\theta_1,\theta_2,\theta_3\: \text{fixed}}} \frac{R_1^0}{\frac{1}{2}\log \text{SNR}_1}= (1-\theta_2)^+
\end{equation}
and
\begin{equation}
d_2^0 = \lim_{\substack{\text{SNR}_1,\text{SNR}_2,\text{INR}_1,\text{INR}_2\rightarrow \infty\\\theta_1,\theta_2,\theta_3\: \text{fixed}}} \frac{R_2^0}{\frac{1}{2}\log \text{SNR}_2}= (1-\frac{\theta_3}{\theta_1})^+
\end{equation}

The problem of obtaining a fair pair of g.d.o.f. can be formulated as a bargaining problem with the feasible set being $\mathcal{D}(\theta_1,\theta_2,\theta_3)$ and the disagreement point being $\mathbf{d}^0$. The two-phase mechanism of coordination proposed in Section IV can also be applied here. In the following, Proposition \ref{thm:gdof1} determines whether the two users have incentives to coordinate in phase 1 and Proposition \ref{thm:gdof2} then solves the bargaining problem in the second phase by selecting the NBS as the desired operating point. A dynamic AOBG can also be formulated for the associated bargaining problem (if the regularity condition holds) but will be omitted here.

{\proposition For the two-user Gaussian IC, the pre-bargaining phase is successful and both
users have incentives to employ an H-K scheme provided one of the following conditions hold. The conditions also list the H-K scheme employed by the users.
\begin{itemize}
\item Strong interference ($\theta_2 \geq \theta_1$ and $\theta_3 \geq 1$): Users always employ HK(0,0);
\item Weak interference ($\theta_2 < \theta_1$ and $\theta_3 < 1$): Users employ HK($1/\text{INR}_2$,$1/\text{INR}_1$) iff $(\theta_1,\theta_2,\theta_3)$ are such that $(d_1^0,d_2^0)$ satisfy (\ref{eqn:weakdof1})-(\ref{eqn:weakdof3}) all with strict inequality;
\item Mixed interference ($\theta_2 \geq \theta_1$ and $\theta_3 < 1$): Users always employ HK($1/\text{INR}_2$,0).
\end{itemize}
\label{thm:gdof1}
}

\vspace{0.4cm}
Unlike in the strong and mixed interference regimes, in the weak interference regime the two users may not both necessarily have the incentives to cooperate. For instance, if $\theta_1 = 1$, $0<\theta_2 \leq \frac{1}{2}$ and $0<\theta_3 \leq \frac{1}{2}$, then $d_1^0+d_2^0 = \varphi_2 = 2-\theta_2-\theta_3 $. In this case, $\mathbf{d}^0$ lies on the boundary of $\mathcal{D}_2$ and is Pareto optimal, thus there is no bargaining outcome that can improve one user's g.d.o.f. without decreasing the other's. Also recall that in Section IV, in the mixed interference regime $a\geq1$ and $b<1$, for finite power constraints $P_1$ and $P_2$, even when $\text{INR}_2 = bP_1>1$, the disagreement point $\mathbf{R}^0$ may not lie strictly inside the rate region achieved by $\text{HK}(1/(bP_1),0)$ and thus pre-bargaining in phase 1 could fail. However, by Proposition \ref{thm:gdof1}, at high SNR's, the pre-bargaining phase is always successful and both users have incentives to employ the simple H-K scheme.

{\proposition Provided that the pre-bargaining in phase 1 is successful, the NBS in phase 2 can be characterized as follows:
\begin{itemize}
\item Strong interference ($\theta_2 \geq \theta_1$ and $\theta_3 \geq 1$): there exists a unique NBS for the bargaining problem $(\mathcal{D}_1,\mathbf{d}^0)$, which is characterized in Proposition 1 with $\mathcal{G} = \mathcal{D}_1$, $\mathbf{g}^0 = \mathbf{d}^0$, $\mathbf{g}^1 = (1\;1)^t$, $\mathbf{A} = (1\; \theta_1)$, $\mathbf{B} = \varphi_1$;
\item Weak interference ($\theta_2 < \theta_1$ and $\theta_3 < 1$): if the bargaining problem $(\mathcal{D}_2,\mathbf{d}^0)$ is essential, there exists a unique NBS which is characterized in Proposition 1 with $\mathcal{G} = \mathcal{D}_2$, $\mathbf{g}^0 = \mathbf{d}^0$, $\mathbf{g}^1 = (1\;1)^t$, $\mathbf{B} = (\varphi_2\;\varphi_3\;\varphi_4)^t$ and
\begin{equation}
\mathbf{A} = \left(
\begin{array}{c c c}
1 & 2 & 1\\
 \theta_1 & \theta_1 & 2\theta_1
\end{array}
\right)^t;
\end{equation}
\item Mixed interference ($\theta_2 \geq \theta_1$) and $\theta_3 < 1$: there exists a unique NBS for the bargaining problem $(\mathcal{D}_3,\mathbf{d}^0)$ which is characterized in Proposition 1 with $\mathcal{G} = \mathcal{D}_3$, $\mathbf{g}^0 = \mathbf{d}^0$, $\mathbf{g}^1 = (1\;1)^t$, $\mathbf{B} = (\varphi_5\;\varphi_6)^t$ and
\begin{equation}
\mathbf{A} = \left(
\begin{array}{c c}
1 & 1\\
 \theta_1 & 2\theta_1
\end{array}
\right)^t.
\end{equation}
\end{itemize}
\label{thm:gdof2}
}

\vspace{0.4cm}
The optimal g.d.o.f. region, the disagreement point and the NBS obtained are illustrated in Fig. \ref{fig:dofregions} for an example in the mixed interference regime. For comparison, we also included the g.d.o.f. region that can be achieved when TDM is used and the corresponding NBS. The g.d.o.f. region in the TDM case is given by $\mathcal{D}_4 = \{\mathbf{d}|\mathbf{d}\geq 0,\; d_1+d_2 \leq 1\}$, which is strictly suboptimal except for some special cases such as the strong interference case with $\theta_1 = \theta_2 = \theta_3 = 1$ and the weak interference case with $\theta_1 = 1$ and $\theta_2 = \theta_3 = \frac{1}{2}$. The TDM NBS is computed as the solution to the bargaining problem $(\mathcal{D}_4, \mathbf{d}^0)$. It can be observed in Fig. \ref{fig:dofregions} that the H-K NBS strictly dominates the TDM NBS. This implies that unlike Fig. \ref{fig:aobgtdm} in Section IV, in certain high SNR regimes, both users would prefer to cooperate using the H-K scheme, rather than the TDM scheme.

\section{Conclusions}
In this paper, we investigated the two-user Gaussian IC, under the assumption that the two users are selfish and interested in coordinating their transmission strategies only when they have incentives to do so. We proposed a two-phase mechanism for the users to coordinate, which consists of choosing a simple H-K type scheme with Gaussian codebooks and fixed power split in phase 1 and bargaining over the achievable rate region (or g.d.o.f. region) to obtain a fair operating point in phase 2. Both the NBS and the dynamic AOBG are considered to solve the bargaining problem in phase 2. As a problem of independent interest, and also as a tool for developing the optimal solution in the strong interference regime, we first studied the MAC before moving on to the IC. We showed that the proposed mechanism can gain substantial rate improvements for both users compared with the uncoordinated case. The results from the dynamic AOBG show that the bargaining game has a unique perfect equilibrium and the agreement is reached immediately in the first bargaining round provided that the associated bargaining problem is regular. The exogenous probabilities of breakdown and which user makes a proposal first also play important roles in the final outcome. When the selfish users' cost of delay in bargaining are not negligible, that is, exogenous probabilities of breakdown are high, the equilibrium outcome deviates from the NBS. We conclude that when we consider coordination and bargaining over the IC, factors such as the users' cost of delay in bargaining and the environment in which bargaining takes place should also be taken into consideration.

In this paper, we derived the cost of delay in bargaining from an exogenous probability of breakdown motivated by the fact that other users in the environment may randomly interrupt the process and the bargaining between a pair of users may terminate in disagreement if no offer is accepted after each round. It would be also interesting to model users' cost of delay in bargaining under other assumptions such as each user's payoff is discounted by a factor of $\delta$ after each round \cite{references:Rubinstein82}\cite{references:Binmore86} or the amount of communication overhead incurred. Finally, the bargaining framework in this paper can be extended to the two-user MIMO IC using the results of \cite{references:Vishwanath04}\cite{references:Shang09_archive}\cite{references:Hsiang08}\cite{references:Telatar07}.

\appendices

\section{The Extensive Game with Perfect Information and Chance Moves}
{\definition An extensive game with perfect information $\Gamma = \langle N,H,P,(\succeq_i)\rangle$ has the following components \cite{references:Martin}:
\begin{itemize}
\item A player set $N$.
\item A history set $H$. Each history in $H$ is a sequence of the form $(e_1,...,e_K)$, where $e_k\;(k=1,...,K)$ is an action taken by a player. If $K$ is $\infty$, the history is infinite. A history $(e_1,...,e_K)$ is \emph{terminal} if it is infinite or if there is no $e_{K+1}$ such that $(e_1,...,e_{K+1})\in H$. The set of terminal histories and that of nonterminal histories are denoted $Q$ and $H\setminus Q$ respectively.
\item A player function $P(h)$ that assigns to each nonterminal history $h\in H\setminus Q$ a member of $N$.
\item For each player $i\in N$ a preference relation $\succeq_i$ on $Q$.
\end{itemize}
}
Let $h$ be a history of length $K$ and $e$ be an action. We denote by $(h,e)$ the history of length $K+1$ consisting of $h$ followed by $e$. After any nonterminal history $h\in H\setminus Q$, player $P(h)$ chooses an action from the set $E(h) = \{e|(h,e)\in H\}$.
{\definition A \emph{strategy} of player $i\in N$ in the extensive game $\Gamma = \langle N,H,P,(\succeq_i)\rangle$ is a function $s_i$ that assigns an action in $E(h)$ to each nonterminal history $h\in H\setminus Q$ for which $P(h) = i$.
}

Let $s = (s_i)_{i\in N}$ be the strategy profile and $s_{-i}$ be the list of strategies $(s_j)_{j\in N\setminus\{i\}}$ for all players except $i$. Given a list $(s_j)_{j\in N\setminus\{i\}}$ and a strategy $s_i$, we also denote by $(s_{-i},s_i)$ the strategy profile. For each strategy profile $s$, we define the \emph{outcome} $O(s)$ of $s$ to be the terminal history that results when each player $i\in N$ follows the precepts of $s_i$.
{\definition
A \emph{Nash equilibrium} of the extensive game $\Gamma = \langle N,H,P,(\succeq_i)\rangle$ is a strategy profile $s^*$ such that for every player $i\in N$ and for every strategy $s_i$ of player $i$, we have
\begin{equation}
O(s^*_{-i},s^*_i)\succeq_i O(s^*_{-i},s_i)
\end{equation}
}
{\definition
The \emph{subgame} of the extensive game $\Gamma = \langle N,H,P,(\succeq_i)\rangle$ that follows the history $h$ is the extensive game $\Gamma(h) = \langle N,H|_h,P|_h,(\succeq_i|_h)\rangle$, where $H|_h$ is the set of sequences $h'$ of actions for which $(h,h')\in H$, $P|_h$ is defined by $P|_h(h') = P(h,h')$ for each $h'\in H|_h$, and $\succeq_i|_h$ is defined by $h'\succeq_i|_h h''$ if and only if $(h,h')\succeq_i (h,h'')$.
}

Given a strategy $s_i$ of player $i$ and a history $h$ in the extensive game $\Gamma$, denote by $s_i|_h$ the strategy that $s_i$ induces in the subgame $\Gamma(h)$ (i.e., $s_i|_h(h') = s_i(h,h')$ for each $h'\in H|_h)$.

{\definition A \emph{subgame perfect equilibrium} of an extensive game $\Gamma = \langle N,H,P,(\succeq_i)\rangle$ is a strategy profile $s^*$ in $\Gamma$ such that for any history $h$, the strategy profile $s^*|_h$ is a Nash equilibrium of the subgame $\Gamma(h)$.
}

A subgame-perfect equilibrium is a Nash equilibrium of the whole game with additional property that the equilibrium strategies induce a Nash equilibrium in every subgame as well.

If there is some \emph{exogenous uncertainty}, the game becomes one with \emph{chance moves} and we denote it by $\langle N,H,P,f_c, (\succeq_i)\rangle$. Under such an extension, $P$ is a function from the nonterminal histories in $H$ to $N \cup \{c\}$ (If $P(h) = c$, then chance determines the action taken after history $h$); for each $h\in H$ with $P(h) = c$, $f_c(\cdot|h)$ is a probability measure on the set $E(h)$ after history $h$;  for each player $i\in N$, $\succeq_i$ is a preference relation on lotteries over the set of terminal histories. The outcome of a strategy profile is a probability distribution over terminal histories and the definition of an subgame perfect equilibrium remains the same as before.

\section{Proof of Proposition \ref{thm:incentive}}
\begin{itemize}
\item
In the strong interference case $a \geq 1$ and $b \geq 1$, we choose optimal $\alpha = \beta = 0$. Treating interference as noise is suboptimal and  $\mathbf{R}^0$ always lies inside $\mathcal{F}$. The bargaining problem $(\mathcal{F},\mathbf{R^0})$ is essential and hence both users always have incentives to cooperate.

\item
In the weak interference $a <1$ and $b<1$, we choose the near-optimal power splits $\alpha = \min(1/(bP_1),1)$ and $\beta = \min(1/(aP_2),1)$. If $bP_1 \leq 1$, the scheme $\text{HK}(1,\beta)$ will not improve user 2's rate over $\mathbf{R}^0$ and hence user 2 does not have an incentive to cooperate using such a scheme. The same will occur to user 1 if $aP_2 \leq 1$ and $\text{HK}(\alpha,1)$ is employed. However, if $aP_2>1$ and $bP_1 > 1$ and $\mathcal{F}\cap \{\mathbf{R}>\mathbf{R}^0\}$ is nonempty when $\alpha = 1/(bP_1)$ and $\beta = 1/(aP_2)$, both users' rates can be improved compared with those in $\mathbf{R}^0$.

\item
In the mixed interference with $a < 1$ and $b \geq 1$, we choose the near-optimal power splits $\alpha = 0$ and $\beta = \min(1/(aP_2),1)$. Similar to the weak case, only if $aP_2>1$ and $\mathcal{F}\cap \{\mathbf{R}>\mathbf{R}^0\}$ is nonempty when $\alpha = 0$ and $\beta = 1/(aP_2)$, it is possible to improve both users' rates relative to those in $\mathbf{R}^0$. Otherwise, at least one user does not have an incentive to cooperate and coordination breaks down.
\end{itemize}

\section{Proof of Proposition \ref{thm:regularity}}
\begin{itemize}
\item
In the strong interference case, at phase 1, the users choose optimal $\alpha = \beta = 0$. The resulting capacity region is shown in Fig. \ref{fig:strreg}. Note that only two extreme points of the region are in the first quadrant and they are $r_1 = (\phi_6-C(P_2),C(P_2))$ and $r_2 = (C(P_1),\phi_6-C(P_1))$. It is easy to show that $R_1^0\leq \phi_6-C(P_2)$ and $R_2^0 \leq \phi_6-C(P_1)$ with equalities holding only when $a = b = 1$. In order for the individual rational efficient frontier to be strictly monotone, it must contain no horizonal or vertical line segments, which requires $R_1^0\geq \phi_6-C(P_2)$ and $R_2^0 \geq \phi_6-C(P_1)$. Hence, the associated bargaining problem is regular iff $a = b = 1$.

\item
In the weak interference case $a<1$ and $b<1$, by Proposition \ref{thm:incentive}, in phase 1, both users have incentives to cooperate using $\text{HK}(1/(bP_1),1/(aP_2))$ if $aP_2>1$, $bP_1 > 1$ and $\mathcal{F}\cap \{\mathbf{R}>\mathbf{R}^0\}$ is nonempty when $\alpha = 1/(bP_1)$ and $\beta = 1/(aP_2)$. The shape of achievable rate region is shown in Fig. \ref{fig:weakreg}. It has been proved in \cite{references:Khandani09} that the points $r_i'\notin\mathcal{F}$ for $i\in \{1,2,...,6\}$. Therefore there are at most\footnote{In \cite{references:Khandani09}, the authors concluded that there should be exactly four extreme points in the first quadrant, but we find that under some parameters one or two of the four points may actually not lie in the first quadrant. For instance, it is possible that $\phi_5-2\phi_2<0$, in which case $r_4$ is not in the first quadrant.} four extreme points in the first quadrant of Fig. \ref{fig:weakreg}, given by
\begin{align}
&r_1 = (\phi_1,\phi_4-2\phi_1)\\
&r_2 = (\phi_4-\phi_3,2\phi_3-\phi_4)\\
&r_3 = (2\phi_3-\phi_5,\phi_5-\phi_3)\\
&r_4 = (\phi_5-2\phi_2,\phi_2)
\end{align}
where $\phi_i,\: i\in\{1,2,...,5\}$ are given in (\ref{eqn:reg1})-(\ref{eqn:reg5}) with $\alpha = 1/(bP_1)$ and $\beta = 1/(aP_2)$.
In order for the individual rational efficient frontier to be strictly monotone, it must contain no horizonal or vertical line segments. If $r_1$ is in the first quadrant, $R_2^0\geq \phi_4-2\phi_1$ must hold and similarly if $r_4$ is in the first quadrant, $R_1^0\geq \phi_5-2\phi_2$ must hold.
Hence, the associated bargaining problem in the weak interference case is regular iff two additional conditions $R_1^0 \geq (\phi_5-2\phi_2)^+$ and $R_2^0 \geq (\phi_4-2\phi_1)^+$ are satisfied.

\item
In the mixed interference case $a<1$ and $b\geq 1$, by Proposition \ref{thm:incentive}, in phase 1, both users cooperate using $\text{HK}(0,1/(aP_2))$ if $aP_2>1$ and $\mathcal{F}\cap \{\mathbf{R}>\mathbf{R}^0\}$ is nonempty when $\alpha = 0$ and $\beta = 1/(aP_2)$. Similar to the weak interference case, there are at most four extreme points in the first quadrant of Fig. \ref{fig:weakreg} except that $r_1' = (\phi_1,\phi_3-\phi_1)$ or $r_5'= (\phi_3-\phi_2,\phi_2)$ may become an extreme point of $\mathcal{F}$, depending on whether the constraint (\ref{eqn:reg4}) or (\ref{eqn:reg5}) is redundant or not respectively. In order for the individual rational efficient frontier to be strictly monotone, it must contain no horizonal or vertical line segments. If $r_1$ and $r_1'$ are both in the first quadrant, $R_2^0\geq \min(\phi_4-2\phi_1,\phi_3-\phi_1)$ must hold and if $r_4$ and $r_5'$ are both in the first quadrant, $R_1^0\geq \min(\phi_5-2\phi_2,\phi_3-\phi_2)$ must hold. Hence, the associated bargaining problem in the mixed interference case $a<1$ and $b\geq 1$ is regular iff two additional conditions $R_1^0 \geq (\min(\phi_5-2\phi_2,\phi_3-\phi_2))^+$ and $R_2^0 \geq (\min(\phi_4-2\phi_1,\phi_3-\phi_1))^+$ are satisfied.
\end{itemize}

\section{Proof of Proposition \ref{thm:gdof1}}
For all interference regimes, since we assume $\theta_i>0$ for $i = 1,2,3$, it immediately follows that $d_1^0< 1$ and $d_2^0< 1$.
\begin{itemize}
\item
In the strong interference regime, we have $\theta_2 \geq \theta_1$ and $\theta_3 \geq 1$. Depending on the values of $\theta_1$, $\theta_2$ and $\theta_3$, we have the following four cases:
\begin{itemize}
\item Case 1: $\theta_2 \geq 1$ and $\theta_3 \geq \theta_1$. In this case, $d_1^0 + \theta_1 d_2^0 = 0$ and $\varphi_1 = \min(\theta_2,\theta_3)>0$. Therefore, $d_1^0 + \theta_1 d_2^0 < \varphi_1$ holds.
\item Case 2: $\theta_2 < 1$ and $\theta_3 \geq \theta_1$. In this case, $d_1^0 + \theta_1 d_2^0 =  1-\theta_2$ and $\varphi_1 = \min(1,\theta_3) = 1> 1-\theta_2$. Therefore, $d_1^0 + \theta_1 d_2^0 < \varphi_1$ holds.
\item Case 3: $\theta_2 \geq 1$ and $\theta_3 < \theta_1$. In this case, $d_1^0 + \theta_1 d_2^0 = \theta_1-\theta_3$ and $\varphi_1 = \min(\theta_2,\theta_1) = \theta_1>\theta_1-\theta_3$. Therefore, $d_1^0 + \theta_1 d_2^0 < \varphi_1$ holds.
\item Case 4: $\theta_2 < 1$ and $\theta_3 < \theta_1$. In this case, $d_1^0 + \theta_1 d_2^0 = 1-\theta_2+\theta_1-\theta_3$ and $\varphi_1 = \min(1,\theta_1)$. Since $1-\theta_2+\theta_1-\theta_3\leq 1-\theta_3<1$ and $1-\theta_2+\theta_1-\theta_3 \leq -\theta_2+\theta_1< \theta_1$, it follows that $1-\theta_2+\theta_1-\theta_3 < \min(1,\theta_1)$. Therefore, $d_1^0 + \theta_1 d_2^0 < \varphi_1$ holds.
\end{itemize}
Hence, $d_1^0 + \theta_1 d_2^0 < \varphi_1$ holds for all the values of the parameters in the range, and we can conclude that $\mathbf{d}^0$ is strictly inside of $\mathcal{D}_1$ and the bargaining problem $(\mathcal{D}_1,\mathbf{d}^0)$ is always essential.
\item
In the weak interference regime, we have $\theta_2 < \theta_1$ and $\theta_3 < 1$. In order for both users to have incentives to cooperate using HK($1/\text{INR}_2$,$1/\text{INR}_1$), $\mathbf{d}^0$ needs to lie strictly inside of $\mathcal{D}_2$, which is not true for all parameters of $\theta_1$,$\theta_2$,$\theta_3$. It happens only when $(\theta_1,\theta_2,\theta_3)$ are such that $(d_1^0,d_2^0)$ satisfy (\ref{eqn:weakdof1})-(\ref{eqn:weakdof3}) all with strict inequality.
\item
In the mixed interference regime, we have $\theta_2 \geq \theta_1$ and $\theta_3 < 1$. Depending on the values of $\theta_1$, $\theta_2$ and $\theta_3$, we have the following four cases:
\begin{itemize}
\item Case 1: $\theta_2 \leq 1$ and $\theta_3 \geq \theta_1$. In this case, $d_1^0 + \theta_1 d_2^0 =d_1^0 + 2\theta_1 d_2^0 = 1-\theta_2$ and $\varphi_5 = 1>1-\theta_2$. Therefore, $d_1^0 + \theta_1 d_2^0 < \varphi_5$ holds. Note that $\varphi_6 = \max(\theta_2+\theta_3,1)\geq 1>1-\theta_2$, hence  $d_1^0 + 2\theta_1 d_2^0 < \varphi_6$ also holds.
\item Case 2: $\theta_2 > 1$ and $\theta_3 \geq \theta_1$. In this case, $d_1^0 + \theta_1 d_2^0 =d_1^0 + 2\theta_1 d_2^0 = 0$ and $\varphi_5 = \theta_2>0$. Therefore, $d_1^0 + \theta_1 d_2^0 < \varphi_5$ holds. Also $\varphi_6 = \theta_2+\theta_3> 0$ and $d_1^0 + 2\theta_1 d_2^0 < \varphi_6$ holds.
\item Case 3: $\theta_2 > 1$ and $\theta_3 < \theta_1$. In this case, $d_1^0 + \theta_1 d_2^0 = \theta_1-\theta_3$ and $\varphi_5 = \min(1+\theta_1-\theta_3,\theta_2)$. Since $\theta_1-\theta_3 < \theta_1 \leq \theta_2$, it follows that $d_1^0 + \theta_1 d_2^0 < \varphi_5$.  $d_1^0 + 2\theta_1 d_2^0 = 2(\theta_1-\theta_3)$ and $\varphi_6 = \theta_1+\theta_2 \geq 2\theta_1 >2(\theta_1-\theta_3)$. Therefore  $d_1^0 + 2\theta_1 d_2^0 < \varphi_6$ also holds.
\item Case 4: $\theta_2 \leq 1$ and $\theta_3 < \theta_1$. In this case, $d_1^0 + \theta_1 d_2^0 = 1-\theta_2+\theta_1-\theta_3\leq 1-\theta_3<1$ and $\varphi_5 = 1$. It follows that $d_1^0 + \theta_1 d_2^0 < \varphi_5$.  $d_1^0 + 2\theta_1 d_2^0 = 1-\theta_2+2(\theta_1-\theta_3)$ and $\varphi_6 = \theta_1+\max(\theta_2,1-\theta_3)$. If $\theta_2\leq 1-\theta_3$, then $\varphi_6 = \theta_1+1-\theta_3$. $d_1^0 + 2\theta_1 d_2^0 - \varphi_6 = -\theta_2+\theta_1-\theta_3\leq -\theta_3<0$. Otherwise if $\theta_2> 1-\theta_3$, then $\varphi_6 = \theta_1+\theta_2$. $d_1^0 + 2\theta_1 d_2^0 - \varphi_6 = 1+\theta_1-2\theta_2-2\theta_3 = -\theta_3 +(1-\theta_3-\theta_2)+(\theta_1-\theta_2)<0$. Therefore $d_1^0 + 2\theta_1 d_2^0 < \varphi_6$ also holds.
\end{itemize}
Hence, $d_1^0 + \theta_1 d_2^0 < \varphi_5$ and $d_1^0 + 2\theta_1 d_2^0 < \varphi_6$ hold for all values of the parameters in the range, and we can conclude that $\mathbf{d}^0$ is strictly inside of $\mathcal{D}_3$ and the bargaining problem $(\mathcal{D}_3,\mathbf{d}^0)$ is always essential.
\end{itemize}

\nocite{*}
\bibliographystyle{IEEEtran}
\bibliography{IEEEabrv,references}

\begin{figure}
\centering
\includegraphics[width = 4in]{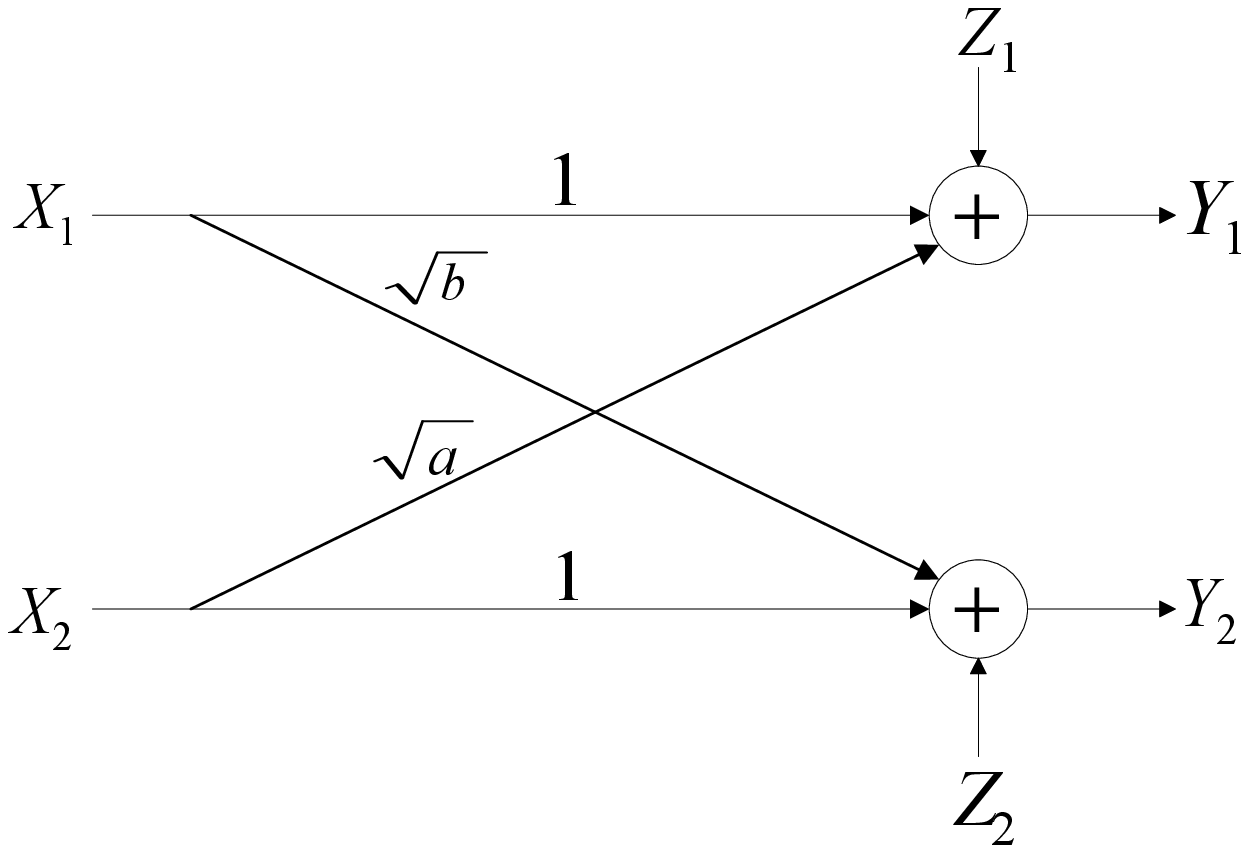}
\caption{Gaussian interference channel}
\end{figure}

\begin{figure}
\centering
\includegraphics[width = 4in]{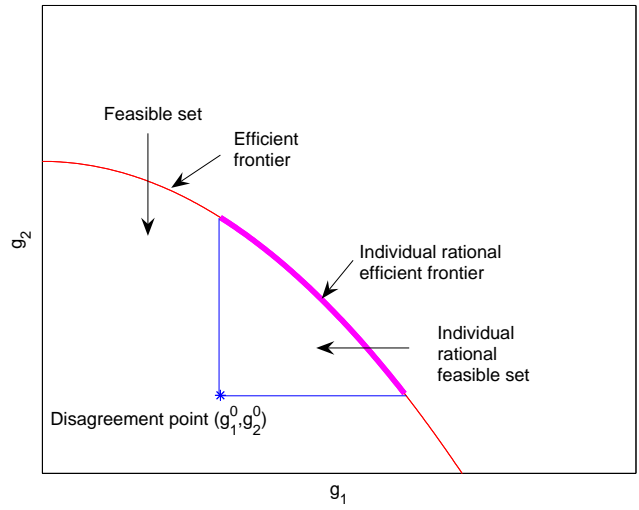}
\caption{Illustration of a bargaining problem.}
\label{fig:frontier}
\end{figure}

\begin{figure}
\centering
\includegraphics[width = 4in]{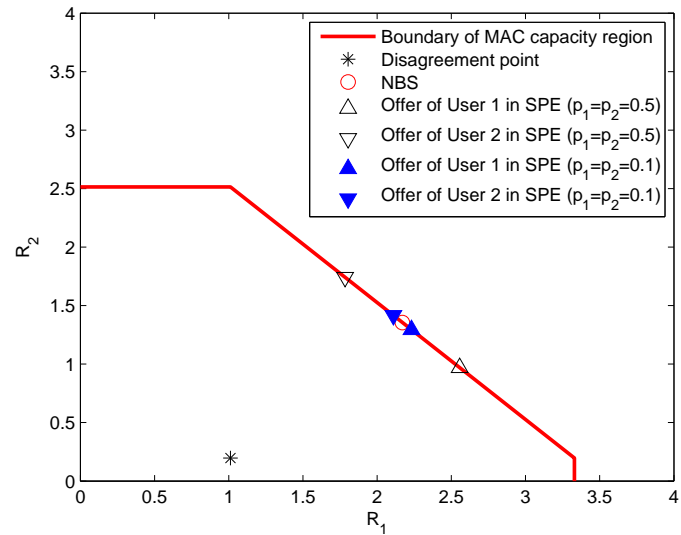}
\caption{Bargaining rates over the MAC when $\text{SNR}_1 = 20$dB, $\text{SNR}_2 = 15$dB}
\label{fig:macnbs}
\end{figure}

\begin{figure*}[ht]
\centering
\subfigure[$\text{SNR}_1 = \text{SNR}_2 = 20$dB]{
\includegraphics[width = 4in]{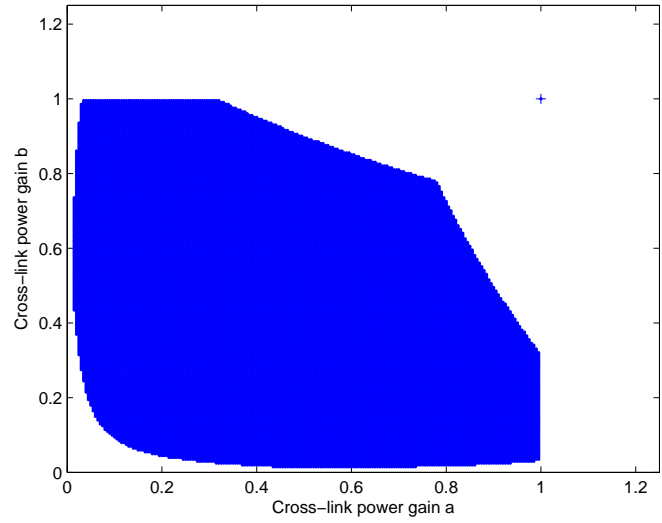}
\label{fig:regular_range1}
}
\subfigure[$\text{SNR}_1 = 20$dB, $\text{SNR}_2 = 30$dB]{
\includegraphics[width = 4in]{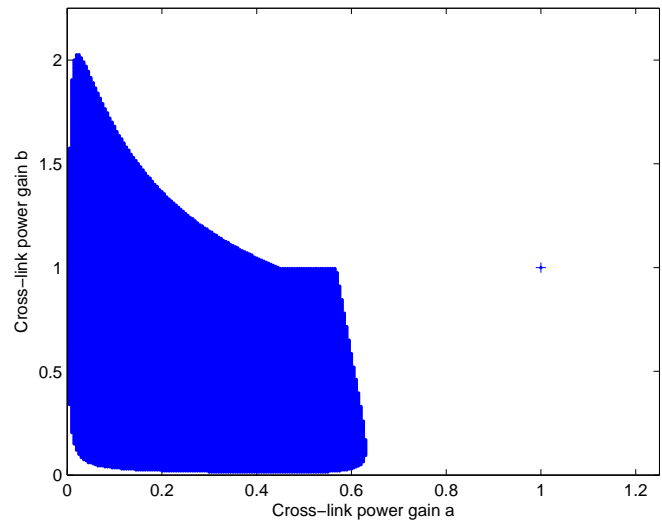}
\label{fig:regular_range2}
}
\caption{Set of cross-link power gains for which pre-bargaining in phase 1 is successful and the associated bargaining problem is regular.}
\label{fig:regular_range}
\end{figure*}

\begin{figure*}[ht]
\centering
\subfigure[Strong interference $a = 3$, $b = 5$]{
\includegraphics[width = 3.8in]{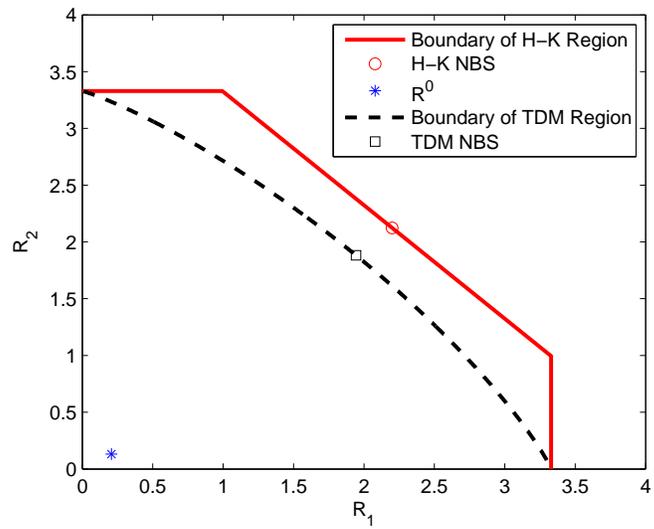}
\label{fig:subfig1}
}
\subfigure[Mixed interference $a = 0.1$, $b = 3$]{
\includegraphics[width = 3.8in]{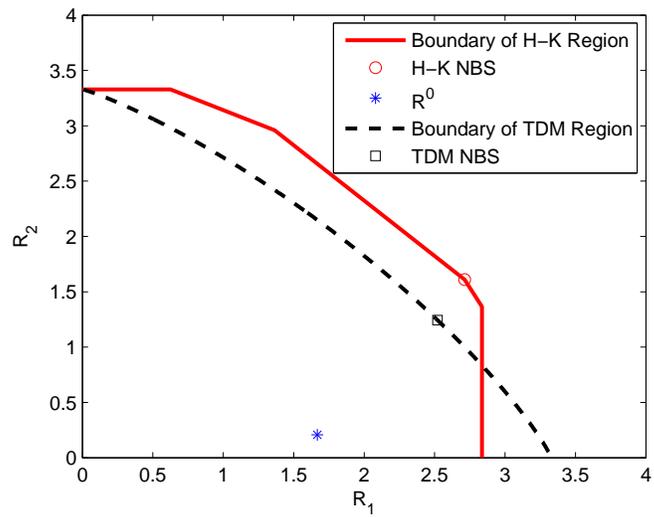}
\label{fig:subfig2}
}
\subfigure[Weak interference $a = 0.2$, $b = 0.5$]{
\includegraphics[width = 3.8in]{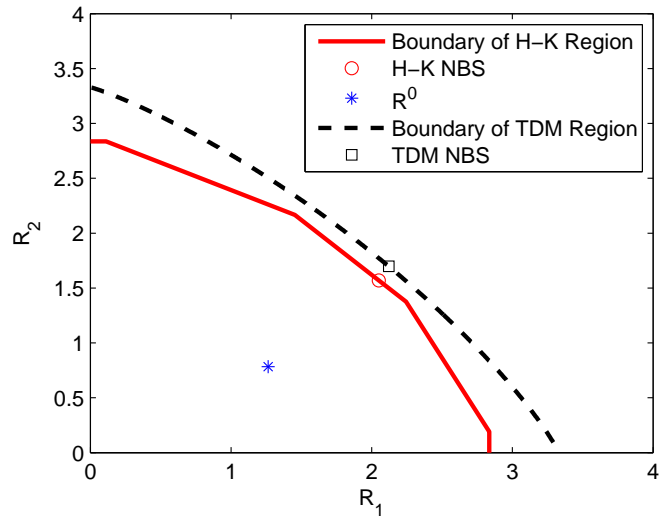}
\label{fig:subfig3}
}
\caption{The H-K and TDM NBS of the Gaussian IC in different interference regimes when $\text{SNR}_1 = \text{SNR}_2 = 20$dB}
\label{fig:hkvstdm}
\end{figure*}
\begin{figure}
\centering
\includegraphics[width = 4in]{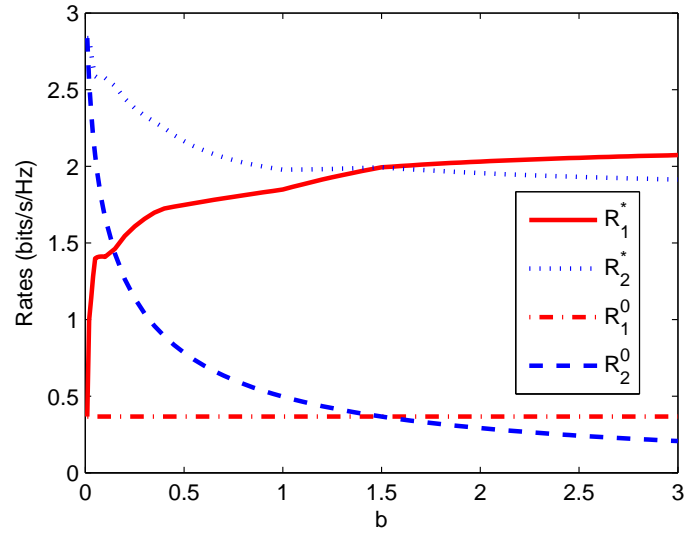}
\caption{Rates in NBS $\mathbf{R}^*$ and disagreement point $\mathbf{R}^0$ as a function of cross-link power gain $b$ when $\text{SNR}_1 = \text{SNR}_2 = 20$dB and cross-link power gain $a = 1.5$.}
\label{fig:ratesvsb}
\end{figure}

\begin{figure}
\centering
\includegraphics[width = 4in]{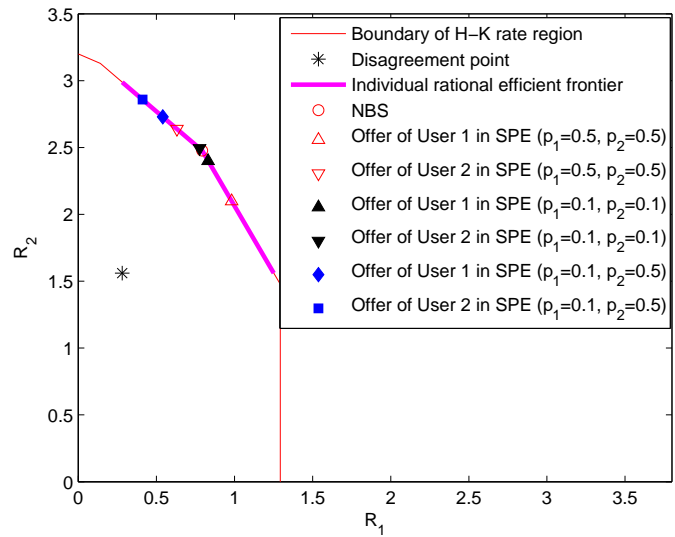}
\caption{The NBS  and perfect equilibrium outcomes of AOBG for IC under mixed interference with $a = 0.2$, $b = 1.2$, $\text{SNR}_1 = 10$dB and $\text{SNR}_2 = 20$dB.}
\label{fig:icaobg}
\end{figure}
\begin{figure}
\centering
\includegraphics[width = 4in]{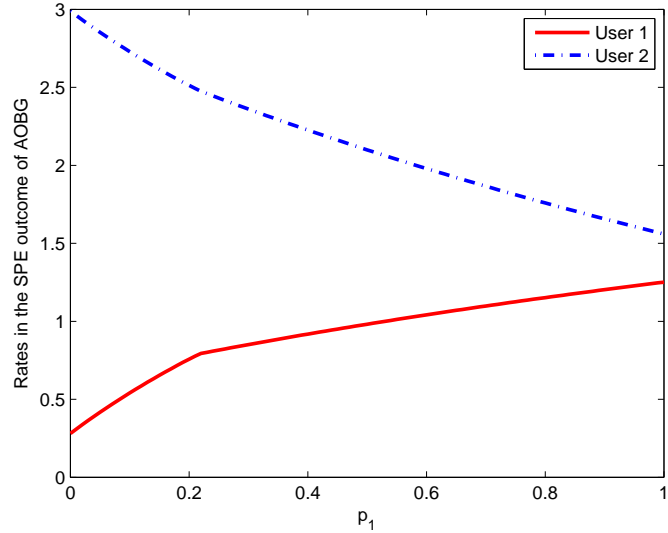}
\caption{Rate of each user in subgame perfect equilibrium of AOBG as a function of probability of breakdown $p_1$ when $p_2 = 0.5$ for IC in mixed interference with $a = 0.2$, $b = 1.2$, $\text{SNR}_1 = 10$dB and $\text{SNR}_2 = 20$dB.}
\label{fig:speratevsp1}
\end{figure}

\begin{figure}
\centering
\includegraphics[width = 4in]{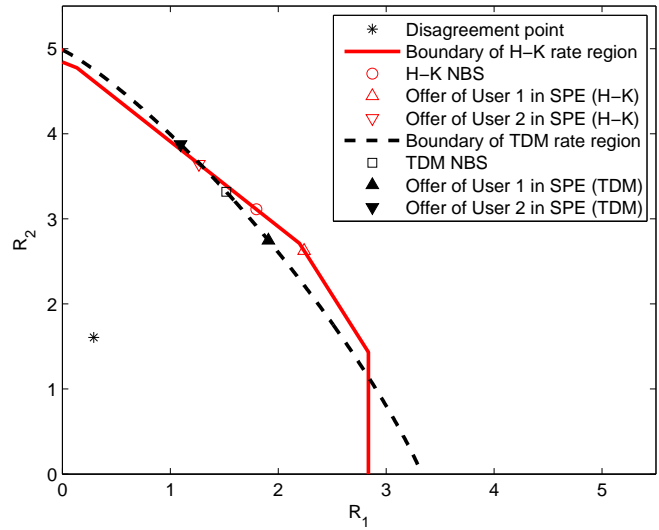}
\caption{Comparison of bargaining outcomes when the H-K scheme and the TDM scheme are used respectively in mixed interference with $\text{SNR}_1 = 20$dB, $\text{SNR}_2 = 30$dB, $a = 0.2$, $b = 1.2$ and $p_1 = p_2 =0.5$.}
\label{fig:aobgtdm}
\end{figure}

\begin{figure}
\centering
\includegraphics[width = 4in]{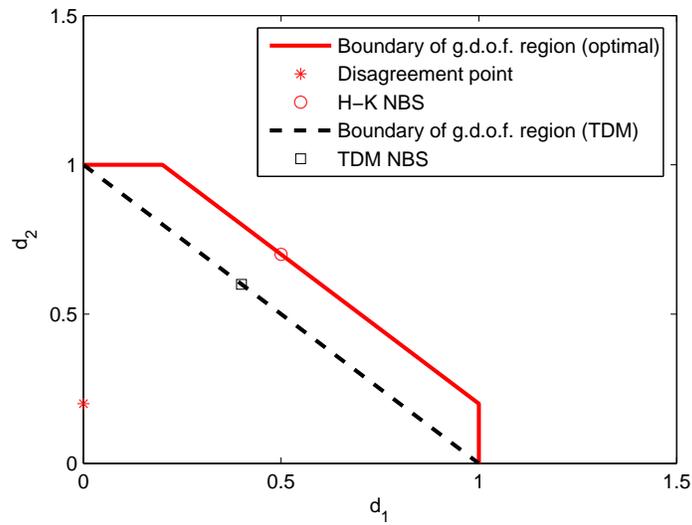}
\caption{The NBS for the g.d.o.f. in mixed interference with $\theta_1 = 1$, $\theta_2 = 1.2$, $\theta_3 = 0.8$}
\label{fig:dofregions}
\end{figure}

\begin{figure*}[ht]
\centering
\subfigure[Strong interference]{
\includegraphics[width = 2.5in]{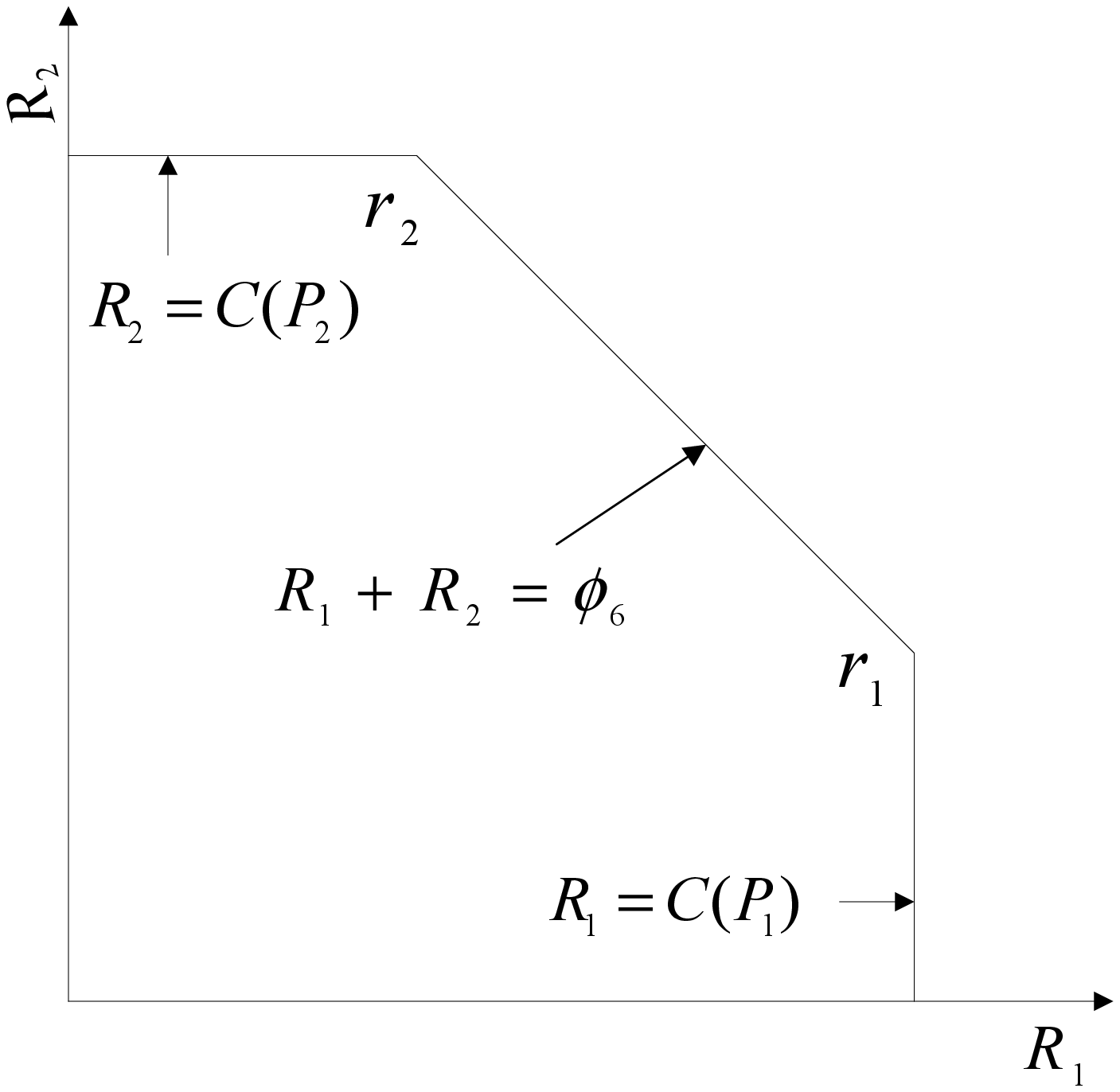}
\label{fig:strreg}
}
\subfigure[Weak or mixed interference]{
\includegraphics[width = 2.5in]{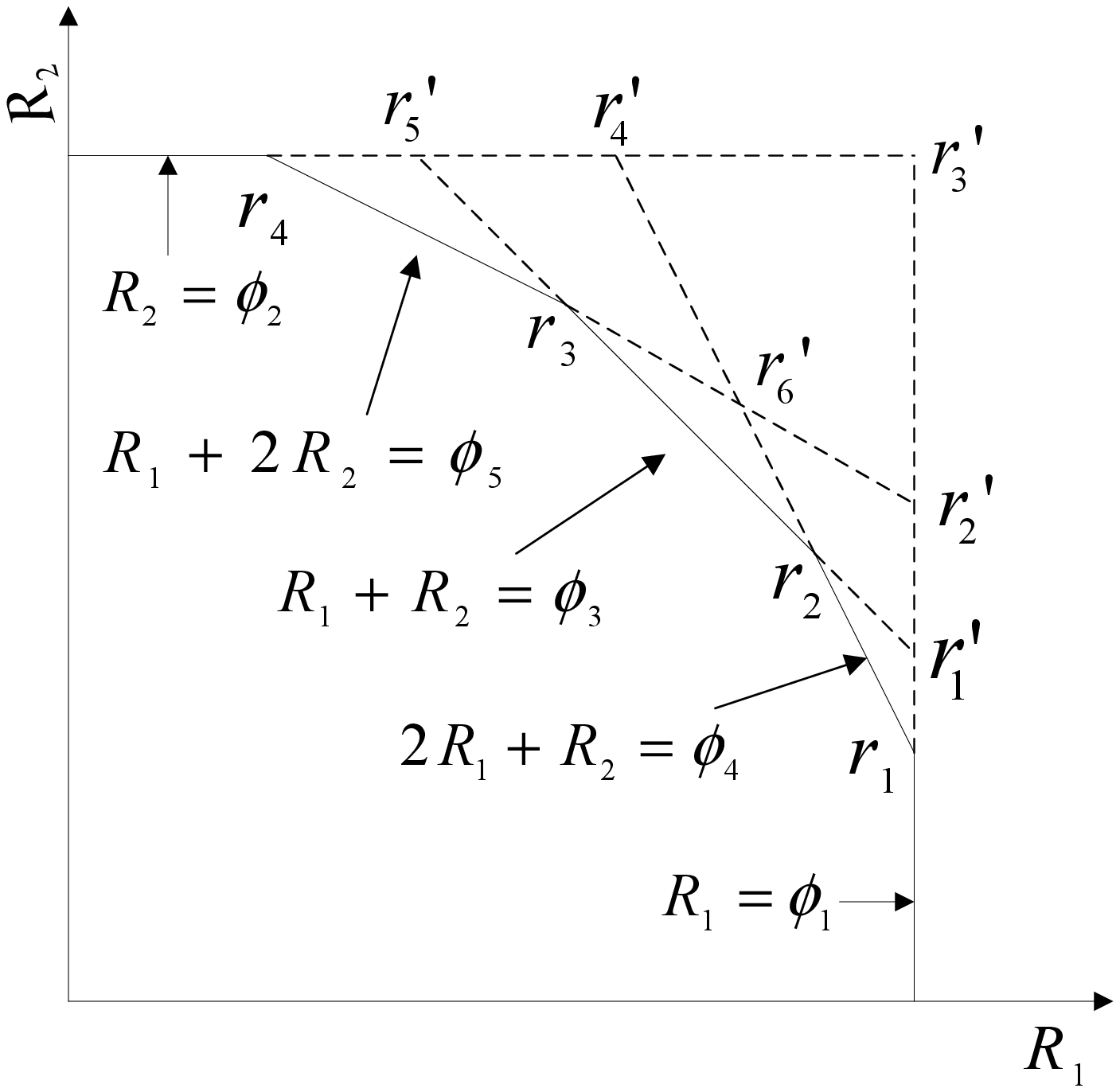}
\label{fig:weakreg}
}
\caption{Achievable rate region using a simple H-K scheme under different interference regimes}
\label{fig:arregions}
\end{figure*}
\end{document}